\documentclass[11pt,a4paper]{article}

\usepackage{jcappub}
\usepackage{epsfig,multicol,bbm}
\usepackage{amsmath}
\usepackage{amsfonts}
\usepackage{amssymb}
\usepackage{url}

\newcommand{\eqreff}[1]{Eq.~\eqref{#1}}
\newcommand{\di}{{\rm d}}

\newcommand{\cs}{c_{\rm s}}

\newcommand{\cb}{c_{\rm b}}
\newcommand{\h}{\mathcal{H}}

\title{Bulk viscous cosmology with causal transport theory}

\author{Oliver~F.~Piattella, J\'{u}lio~C.~Fabris and Winfried~Zimdahl}
\affiliation{Departamento de F\'isica, Universidade Federal do Esp\'irito Santo, Avenida Ferrari 514, 29075-910 Vit\'oria, Esp\'irito Santo, Brazil}

\emailAdd{oliver.piattella@gmail.com}
\emailAdd{fabris@pq.cnpq.br}
\emailAdd{winfried.zimdahl@pq.cnpq.br}

\abstract{We consider cosmological scenarios originating from a single imperfect fluid with bulk viscosity and apply Eckart's and both the full and the truncated M\"{u}ller-Israel-Stewart's theories as descriptions of the non-equilibrium processes. Our principal objective is to investigate if the dynamical properties of Dark Matter and Dark Energy can be described by a single viscous fluid and how such description changes when a causal theory (M\"{u}ller-Israel-Stewart's, both in its full and truncated forms) is taken into account instead of Eckart's non-causal theory. To this purpose, we find numerical solutions for the gravitational potential and compare its behaviour with the corresponding $\Lambda$CDM case. Eckart's and the full causal theory seem to be disfavoured, whereas the truncated theory leads to results similar to those of the $\Lambda$CDM model for a bulk viscous speed in the interval $10^{-11} \ll \cb^2 \lesssim 10^{-8}$. Tentatively relating such value to a square propagation velocity of the order of $T/m$ of perturbations in a non-relativistic gas of particles with mass $m$ at the epoch of matter-radiation equality, this may be compatible with a mass range $0.1 \mbox{ GeV} \lesssim m \ll 100 \mbox{ GeV}$.}


\begin{document}
\maketitle


\section{Introduction}

On large scales our present universe seems to be dynamically dominated by two so far unknown components called Dark Energy (DE) and Dark Matter (DM), at least if the validity of Einstein's General Relativity is taken for granted. To
uncover the physical nature of these components is a major activity within the cosmologists' community. Lacking a fundamental theoretical understanding, the attempts to interpret the wealth of observational data has led to a host of phenomenological models with varying degrees of physical motivation. The most favoured one is the $\Lambda$CDM model, which is also the simplest. Even though this model has good chances to be the ``correct" one, it continues to make sense to study alternatives that may serve to complement our understanding of the cosmological dynamics, let alone the cosmological constant problem \cite{Weinberg:1988cp}.

Given the large variety of phenomenological approaches, it seems desirable to look for unifying aspects of at least some of them. This may also imply a modified view on the status of the $\Lambda$CDM model. For example, the latter can be regarded as a particular case of the generalised Chaplygin gas (GCG) models \cite{Kamenshchik:2001cp, Bento:2002ps, Gorini:2007ta, Piattella:2009da}. In turn, it was shown that the GCG can be interpreted on the basis of interacting holographic DE models \cite{Zimdahl:2007ne, Zimdahl:2007zz}, where the interaction parameter in the dark sector determines the GCG parameters.

A further line of research have been dissipative DE models, in which the negative pressure, responsible for the current acceleration, is an effective bulk viscous pressure, i.e. a non-equilibrium phenomenon.
This parallels related approaches for an inflationary phase in the early universe \cite{Barrow:1988yc, Maartens:1995wt, Zimdahl:1996ka, Maartens:1996vi, Zimdahl:1999tn}. It was argued in \cite{Zimdahl:2000zm, Balakin:2003tk}, that a viscous pressure can play the role of an agent that drives the present acceleration of the Universe. Note that the possibility of a viscosity dominated late epoch of the Universe with accelerated expansion was already mentioned by Padmanabhan and Chitre in 1987 \cite{Padmanabhan:1987dg}. For more recent investigations see, e.g. \cite{Fabris:2005ts, Colistete:2007xi, Szydlowski:2006ma, Avelino:2008ph, Li:2009mf, Avelino:2010pb} and references therein. It has also been pointed out in \cite{Fabris:2005ts, Colistete:2007xi, Szydlowski:2006ma} that there exists a correspondence between dissipative and GCG models such that for appropriate parameters combinations both models result in an identical dynamics.

It is expedient to emphasise, however, that these very different manners of understanding the apparently same dynamics, say of the $\Lambda$CDM model, are restricted to a homogeneous and isotropic background. At the perturbative level, the different mentioned approaches imply a generally different dynamics. The unifying view on the background level is accompanied by a difference in the perturbation dynamics, e.g., as a consequence of fluctuations in the relevant interaction rates \cite{Zimdahl:2007ne, Zimdahl:2007zz}. This circumstance makes the different approaches potentially testable alternatives to the ``standard" $\Lambda$CDM model while sharing the same background dynamics of the latter.

Traditionally, for the description of non-equilibrium thermodynamical processes the theories of Eckart \cite{Eckart:1940te} and Landau and Lifshitz \cite{landau1958fluid} were used. Due to the work of M\"{u}ller \cite{muller1967paradoxon}, Israel \cite{Israel:1976tn}, Israel and Stewart \cite{Israel:1979wp}, Pav\'on, Jou and Casas-V\'azquez \cite{pavón1982covariant}, Hiscock and Lindblom \cite{Hiscock:1983zz} it became clear, however, that Eckart-type theories suffer from serious drawbacks concerning causality and stability. These difficulties could be traced back to their restriction to first-order deviations from equilibrium. If one includes second-order deviations as well, the corresponding problems disappear. Cosmological implications of second-order theories were first considered by Belinskii, Nikomarov and Khalatnikov \cite{belinskii1979investigation}, followed by Pav\'on, Jou and Casas-V\'azquez \cite{pavón1982covariant} and other authors \cite{Maartens:1995wt, Zimdahl:1996ka, Maartens:1996vi}. Reheating and cosmological particle production on the basis of causal thermodynamics have been discussed in \cite{Zimdahl:1996tg, Zimdahl:1999tn}.

One should be aware that discussing the issue of bulk-viscous driven accelerated expansion at all, implies in any case an extrapolation of non-equilibrium thermodynamical theories beyond the range for which their applicability was strictly justified. Bulk viscous inflation, if it exists, is a far-from-equilibrium phenomenon, while even the full causal second-order M\"{u}ller-Israel-Stewart's (MIS) theory holds for small deviations from equilibrium. Therefore, all theoretical conclusions are necessarily tentative but, hopefully, they nevertheless will provide an indication of the correct behaviour far from equilibrium.

The impact of bulk viscosity on the background expansion of the universe is widely investigated in the literature, see e.g. \cite{Murphy:1973zz, belinskii1979investigation, Pavon:1990qf, Chimento:1993zc, zakari1993equations, Chimento:1997vy, mak1998full, DiPrisco:2000dw, Szydlowski:2005uq, Szydlowski:2006ma, Hu:2005fu, Maartens:1995wt, Belinchon:2004zk, Folomeev:2007gr, Tawfik:2009nh}. However, the perturbative analysis of the viscous cosmological models is not as widely addressed, despite its crucial importance. Among those papers which do perform a perturbative analysis, we cite \cite{Fabris:2005ts, Colistete:2007xi, HipolitoRicaldi:2009je, Li:2009mf, HipolitoRicaldi:2010mf}. A common characteristic of these papers is the analysis of a fluid whose equilibrium pressure is negligible with respect to the bulk viscous one. Moreover, another common point is that the dissipation process is described within Eckart's theory. Interestingly, there exist choices of the bulk viscosity parameters which give predictions for the matter power spectrum in agreement with observation \cite{HipolitoRicaldi:2009je,HipolitoRicaldi:2010mf}. On the other hand, the results found in \cite{Li:2009mf} seem to doom the possibility of describing DM and DE together as a bulk viscosity effect because a very strong integrated Sachs-Wolfe effect would be produced.

The present paper relies on a cosmological dynamics, realized by a single imperfect fluid with bulk viscosity and vanishing equilibrium pressure. However, we investigate how the picture changes when taking into account the causal M\"{u}ller-Israel-Stewart theory, both in its full and truncated forms, instead of Eckart's theory. Our focus is on the behaviour of the gravitational potential, which we compare with its $\Lambda$CDM counterpart. Assuming that the results of the standard $\Lambda$CDM model \textit{grosso modo} reproduce the observations, we will discard any viscous model that differs substantially from this standard.

For both Eckart's and the full MIS theories, our results seem to disfavour the possibility of unifying DM and DE via bulk viscosity. Only the truncated case seems to remain a valid option for bulk viscous speeds of sound $\cb^2 \lesssim 10^{-8}$. It may be interesting to mention that extremely small values for the speed of sound are also known from certain scalar-field DE models \cite{Creminelli:2008wc}.

The paper is structured as follows: in Sec.~\ref{sec:BulkDyn} we present the fundamental equations of bulk viscous cosmology. In Sec.~\ref{sec:Perts} we set up the formalism necessary to investigate perturbations in the bulk viscous fluid. In Sec.~\ref{Sec:bgexpansion} we adopt and motivate an assumption for the viscosity pressure and in Sec.~\ref{Sec:Results} we show our results, embodied in the evolution of the gravitational potential for the viscous models compared to the $\Lambda$CDM one. In Sec.~\ref{sec:conclusions} we draw our conclusions. We choose the  $-2$ signature for the metric and $c = 1$ units.


\section{ Dynamics of the viscous fluid}\label{sec:BulkDyn}

A flat, homogeneous and isotropic universe is described by the Robertson-Walker (RW) line element
\begin{equation}\label{FLRWmetric}
ds^{2} = dt^{2} - a(t)^{2}\delta_{ij}dx^idx^j = a(\eta)^{2}\left(\di\eta^{2} - \delta_{ij}dx^idx^j\right)\;,
\end{equation}
where $t$ is the cosmic time and $\eta$ the conformal one. Consider a relativistic fluid with bulk viscosity. We refer the reader to \cite{Maartens:1995wt, Zimdahl:1996ka, Maartens:1996vi} for extensive reviews of the topic. Write the total pressure as $P = p + \Pi$, i.e. consider a splitting into the equilibrium part $p$ plus the bulk viscous pressure contribution $\Pi$. In presence of the latter, the stress-energy tensor has the following form:
\begin{equation}\label{bulkviscT}
 T_{\mu\nu} = \left(\rho + P\right)u_{\mu}u_{\nu} - pg_{\mu\nu}\;,
\end{equation}
where $u_{\mu}$ is the four-velocity and $g_{\mu\nu}$ is given by \eqreff{FLRWmetric}. Hence, the bulk viscous pressure part simply adds up to its equilibrium counterpart.

Neglecting the baryon and the radiation components,\footnote{The baryon component is always subdominant whereas radiation is negligible well after decoupling, which is the timespan we are interested in.} Friedmann's equation and the energy conservation equation read
\begin{equation}\label{Feq}
H^2 = \frac{\h^{2}}{a^2} = \frac{8\pi G}{3}\rho\;, \qquad \dot{\rho} = -3H\left(\rho + P\right)\;,
\end{equation}
where $H = \dot{a}/a$ is the Hubble parameter and $\h \equiv a'/a$. Throughout the paper, the dot denotes derivation with respect to (wrt) the cosmic time whereas the prime denotes the derivation wrt the conformal time. Combining the two equations in \eqref{Feq}, one obtains the acceleration equation
\begin{equation}\label{acceq}
\frac{\dot{H}}{H^{2}} = -\frac{3}{2}\left(1 + \frac{P}{\rho}\right)\;,
\end{equation}
which can be solved for $P/\rho$, resulting in
\begin{equation}\label{wq}
\frac{P}{\rho} = \frac{1}{3} \left(2q - 1\right)\;, \qquad q := - 1 - \frac{\dot H}{H^2}\;,
\end{equation}
where $q$ is the deceleration parameter.

The evolution of $\Pi$ is governed by a transport equation whose form depends on the theory adopted for the dissipative processes. Eckart's theory \cite{Eckart:1940te} is based on the algebraic relation
\begin{equation}\label{Eceq}
 \Pi = -\theta\zeta\;,
\end{equation}
where $\zeta$ is the bulk viscosity coefficient and $\theta \equiv \nabla_{\mu}u^{\mu}$ is the expansion scalar, i.e. the covariant divergence of the velocity field. For the RW metric one has $\theta = 3H = 3\h/a$. On the other hand, in the MIS theory \cite{muller1967paradoxon, Israel:1976tn, Israel:1979wp} the evolution of $\Pi$ is described by the following transport equation:
\begin{equation}\label{ISceq}
\tau\Pi^{\bullet} + \Pi = -\theta\zeta - \frac{1}{2}\tau\Pi\left[\theta - \frac{\left(\zeta/\tau\right)^\bullet}{\left(\zeta/\tau\right)} - \frac{T^\bullet}{T}\right]\;,
\end{equation}
where $\tau$ is a relaxation time and $T$ is the temperature. The bullet $\bullet$ in \eqreff{ISceq} denotes
\begin{equation}\label{dotIS}
\Pi^\bullet := u^{\mu}\nabla_{\mu}\Pi\;,
\end{equation}
i.e. derivation along the fluid wordline. In the RW case it reduces to the derivation wrt the cosmic time, but we prefer to keep a separate notation in order to avoid confusion when considering perturbations of \eqref{ISceq}. Equation~\eqref{ISceq} has often been employed in a truncated form, i.e.
\begin{equation}\label{ISceq2trunc}
 \tau\Pi^\bullet + \Pi = -\theta\zeta\;,
\end{equation}
which we are going to investigate in detail, comparing it with the full theory and Eckart's one.

Note that, assuming the bulk viscous pressure $\Pi$ to be subject to causal thermodynamics does not only
introduce the relaxation time as an additional parameter but also, associated with the relaxation time, a propagation
velocity for viscous perturbations which is different from the adiabatic sound speed.

The temperature term in Eq.~\eqref{ISceq} can be calculated from Gibbs' integrability condition \cite{Maartens:1996vi}:
\begin{equation}\label{Gintcond}
 n\frac{\partial T}{\partial n} + (\rho + p)\frac{\partial T}{\partial \rho} = T\frac{\partial p}{\partial \rho}\;.
\end{equation}
Assuming an equation of state $p = p(\rho, n)$, integration of \eqref{Gintcond}  provides us with a functional form for $T = T(\rho, n)$.  Then, energy and particle number conservations allow us to determine
\begin{equation}\label{Tevoeq}
\frac{T^\bullet}{T} = -\theta\left[\frac{\partial p}{\partial \rho} + \frac{\Pi}{T}\frac{\partial T}{\partial \rho}\right]\;.
\end{equation}
In this paper, we assume the temperature and the equilibrium pressure to be barotropic, i.e. described by $T = T(\rho)$ and $p = p(\rho)$. Defining $\partial p/\partial\rho = dp/d\rho := \cs^2$ as the adiabatic speed of sound, \eqreff{Gintcond} takes the form
\begin{equation}\label{Gintcond2}
 \frac{1}{T}\frac{dT}{d\rho} = \frac{\cs^2}{\rho + p}\;
\end{equation}
and \eqreff{Tevoeq} becomes
\begin{equation}\label{Tevoeq2}
\frac{T^\bullet}{T} = -\theta\cs^2\left(1 + \frac{\Pi}{\rho + p}\right)\;.
\end{equation}
(With our assumptions $T = T(\rho)$ and $p = p(\rho)$ we disregard  a possible microscopic description on the basis of kinetic theory). Under this condition, we can cast the MIS transport equation in the following form:
\begin{equation}\label{ISceq2}
 \tau\Pi^\bullet + \Pi = -\theta\zeta - \frac{1}{2}\tau\Pi\left[\theta - \frac{\left(\zeta/\tau\right)^\bullet}{\left(\zeta/\tau\right)} + \theta\cs^2\left(1 + \frac{\Pi}{\rho + p}\right)\right]\;.
\end{equation}
The coefficients $\tau$ and $\zeta$ are, in general, functions of the time (or of the energy density).  They are not completely arbitrary but related to $\cb^2$, the square of the velocity at which a viscous perturbation propagates \cite{Hiscock:1983zz} (see also \cite{Maartens:1996vi}):
\begin{equation}\label{cb2}
 \cb^2 = \frac{\zeta}{(\rho + p)\tau}\;.
\end{equation}
Remarkably, this propagation speed depends also on the equilibrium pressure contribution. Given that $\cb^2$ has to be added to $\cs^2$, for causality reasons we have
\begin{equation}\label{cb2cs2}
 \cb^2 + \cs^2 \le 1\;,
\end{equation}
which provides a constraint on a possible functional form of $\tau$ and $\zeta$ once the adiabatic speed of sound is given.


\section{Perturbative dynamics of the viscous fluid}\label{sec:Perts}

In this section we provide at first the general perturbation equations which then are specified to Eckart's and MIS theories. We adopt the Bardeen gauge-invariant formalism \cite{Bardeen:1980kt, Mukhanov:1990me, mukhanov2005physical}. See also \cite{Giovannini:2005ii} for an extensive review of Eckart's case.

Consider the following perturbation of the RW metric:
\begin{equation}\label{pertmet}
 ds^2 = a(\eta)^2\left[(1 + 2\Phi)d\eta^2 - (1 - 2\Phi)d{\bf x}^2\right]\;.
\end{equation}
The background velocity field $u_{\mu}$ has components $u_0 = a$ and $u_i = 0$. The perturbed zero-component is $\delta u_0 = a\Phi$, $\delta u^0 = -\Phi/a$. A general perturbation of the stress-energy tensor \eqref{bulkviscT} has the form
\begin{eqnarray}
 \delta T^{0}{}_{0} &=& \delta\rho\;,\\
\delta T^i{}_{0} &=& \left(\rho + p + \Pi\right)v^i\;,\\
\delta T^i{}_{j} &=& -\delta^i{}_{j}\left(\delta p + \delta\Pi\right)\;,
\end{eqnarray}
where we have defined $v^i := a\delta u^i$. At first order, Einstein's equations take the form
\begin{eqnarray}
\label{00EE}\Delta\Phi - 3\h\left(\Phi' + \h\Phi\right) &=& 4\pi Ga^2\delta\rho\;,\\
\label{0iEE}\Delta\left(\Phi' + \h\Phi\right) + \left(\h^2 - \h'\right)\Theta &=& 0\;,\\
\label{ijEE} \Phi'' + 3\h\Phi' + \left(\h^2 + 2\h'\right)\Phi &=& 4\pi Ga^2\left(\delta p + \delta\Pi\right)\;,
\end{eqnarray}
where we have used the divergence of the $(0-i)$ equation and defined $\Theta\equiv\partial_iv^i$. The expression for the perturbed bulk viscous contribution $\delta\Pi$ depends on the choice of the theory.

\subsection{Eckart's theory}

From \eqreff{Eceq}, a perturbation of $\Pi$ may be formally written as
\begin{equation}
 \delta\Pi = -\delta\theta\zeta - \theta\delta\zeta\;.
\end{equation}
With $\theta = \nabla_\mu u^\mu$ we find
\begin{equation}
\delta\theta = \partial_\mu \delta u^\mu + \Gamma^{\mu}_{\rho\mu}\delta u^{\rho} + \delta\Gamma^{\mu}_{\rho\mu}u^{\rho}\;,
\end{equation}
where $\Gamma^{\mu}_{\rho\mu}$ is the Levi-Civita connection of the RW metric and $\delta\Gamma^{\mu}_{\rho\mu}$ is its perturbation. Starting from \eqreff{FLRWmetric} and \eqreff{pertmet} and working out the above expression, it is straightforward to obtain
\begin{equation}\label{deltatheta}
 a\delta\theta = \partial_i v^i - 3\left(\Phi' + \h\Phi\right) = \Theta - 3\left(\Phi' + \h\Phi\right)\;.
\end{equation}
Therefore,
\begin{equation}\label{dPiEckart}
\delta\Pi = - \frac{3\h}{a}\delta\zeta - \frac{\zeta}{a}\left[\Theta - 3\left(\Phi' + \h\Phi\right)\right]\;,
\end{equation}
is the perturbed bulk viscous pressure contribution we are looking for.

\subsection{M\"{u}ller-Israel-Stewart's theory}

In the MIS theory, the bulk viscous pressure contribution evolves according to \eqreff{ISceq2}. In the following we shall always replace $\zeta$ by  $\cb$. Note that this is not possible for Eckart's case since it has $\tau = 0$, so that $\cb^2$ diverges, cf. \eqreff{cb2}.
In order to simplify the notation, let $f = f(\rho) := \zeta/\tau$. This choice also leaves open the possibility of a variable $\cb^2$, although we do not consider this case in the present paper.

Equation~\eqref{ISceq2} can then be cast in the compact form
\begin{equation}\label{MISteviscousratio2}
\Pi^\bullet + \frac{1}{\tau}\Pi = -\theta\left[f(\rho) + \frac{\Pi}{2} g(\rho) + \frac{\Pi^2}{2}h(\rho)\right]\;,
\end{equation}
where we have defined
\begin{equation}\label{ghfunc}
 g(\rho) := 2 + \cs^2(\rho) + \frac{\rho + p(\rho)}{\cb^2(\rho)}\;\frac{d\cb^2(\rho)}{d\rho}\;, \qquad h(\rho) := \frac{1}{\cb^2(\rho)}\frac{d\cb^2(\rho)}{d\rho} + \frac{1 + 2\cs^2(\rho)}{\rho + p(\rho)}\;.
\end{equation}
The derivative term in \eqreff{MISteviscousratio2} is perturbed as follows:
\begin{equation}
 \delta\left(\Pi^\bullet\right) = \delta\left(u^\mu\partial_\mu\Pi\right) = - \frac{\Phi}{a}\Pi' + \frac{1}{a}\delta\Pi'\;.
\end{equation}
The first-order equation that results from \eqreff{MISteviscousratio2} is
\begin{eqnarray}\label{MISteviscousratio2pert}
\frac{1}{a}\delta\Pi' + \delta\left(\frac{\Pi}{\tau}\right) = \frac{\Phi}{a}\Pi' - \delta\theta\left[f(\rho) + \frac{\Pi}{2} g(\rho) + \frac{\Pi^2}{2}h(\rho)\right]\nonumber\\ - \theta\delta\rho\left[\frac{df(\rho)}{d\rho} + \frac{\Pi}{2}\frac{dg(\rho)}{d\rho} + \frac{\Pi^2}{2}\frac{dh(\rho)}{d\rho}\right] - \theta\delta\Pi\left[\frac{g(\rho)}{2} + \Pi h(\rho)\right]\;.
\end{eqnarray}
Substituting $\Pi'/a$ from  the zeroth order of \eqref{MISteviscousratio2}, one obtains
\begin{eqnarray}\label{MISteviscousratio3pert}
\frac{1}{a}\delta\Pi' + \delta\left(\frac{\Pi}{\tau}\right) &=& -\frac{\Phi\Pi}{a} - \theta\Phi\left[f(\rho) + \frac{\Pi}{2} g(\rho) + \frac{\Pi^2}{2}h(\rho)\right] - \delta\theta\left[f(\rho) + \frac{\Pi}{2} g(\rho) + \frac{\Pi^2}{2}h(\rho)\right]\nonumber\\ &-& \theta\delta\rho\left[\frac{df(\rho)}{d\rho} + \frac{\Pi}{2}\frac{dg(\rho)}{d\rho} + \frac{\Pi^2}{2}\frac{dh(\rho)}{d\rho}\right] - \theta\delta\Pi\left[\frac{g(\rho)}{2} + \Pi h(\rho)\right]\;.
\end{eqnarray}
This equation determines the perturbation $\delta\Pi$ that enters the right-hand side of \eqref{ijEE}.

The truncated form of \eqref{MISteviscousratio3pert} can be obtained by putting $g = h = 0$:
\begin{eqnarray}\label{MISteviscousratio2perttrunc}
\frac{1}{a}\delta\Pi' + \delta\left(\frac{\Pi}{\tau}\right) = -\frac{\Phi\Pi}{\tau} - \theta\Phi f(\rho) - \delta\theta f(\rho) - \theta\delta\rho\frac{df(\rho)}{d\rho}\;.
\end{eqnarray}
In the following section we analyse special forms of $\zeta$ and $\tau$ which provide a background expansion similar to the one produced by the generalised Chaplygin gas \cite{Kamenshchik:2001cp, Bento:2002ps}.


\section{Ansatz for the pressure }\label{Sec:bgexpansion}

Up to now, our formalism for a one-component bulk viscous fluid with barotropic equation of state has been quite general. Now and hereafter we also assume the effective pressure to be totally dissipative, i.e. $P = \Pi$. This implies $p = 0$ and $\cs^2 = 0$. Moreover, we demand $\cb^2$ to be a constant, so that in MIS theories we have $\zeta = \cb^2\rho\tau$.

In the causal theory the viscous pressure becomes a dynamical degree of freedom, subject to a first order
differential equations, both in the background and on the level of perturbations. However, in the context of this paper these differential equations are \textit{not} used to determine the viscous pressure and its perturbation. Instead, we shall assume the viscous pressure to be known. Then, the causal evolution equation provides us with a relation between the relaxation time and the Hubble time. This strategy is motivated as follows. Our present Universe is well described by the $\Lambda$CDM model, both at the RW level and for linear perturbations about that background. As far as the former is concerned, the $\Lambda$CDM model corresponds to a constant pressure in a one-component model of the cosmic substratum. In order to admit also deviations from the $\Lambda$CDM model, we start with an ansatz
\begin{equation}
\frac{\Pi}{\rho} = - \mu \left(\frac{\rho}{\rho_{0}}\right)^{-n/2}\;,\label{eoszeroo}
\end{equation}
where $\mu$ is a constant and $\rho_0$ is the present time (i.e. for $a = a_0$) energy density of the viscous fluid. For $n = 2$ we recover the constant pressure case. Equation~\eqref{eoszeroo} represents the equation of state of a generalised Chaplygin gas \cite{Kamenshchik:2001cp, Bento:2002ps}. Consistently, by solving the energy conservation equation one can find, taking $a_{0}=1$,
\begin{equation}
\rho = \rho_{0}\left[\mu + \left(1 - \mu\right)a^{-3n/2}\right]^{2/n}\;.
\label{rhosolve}
\end{equation}
In the limit $a \ll 1$ the solution \eqref{rhosolve} reproduces a matter dominated universe with $\rho \propto a^{-3}$, while in the opposite limit the energy density is similar to that of a cosmological constant. Moreover, for $n = 2$  we recover the $\Lambda$CDM model while for $n = 4$ the expression \eqref{rhosolve} describes the energy density of the ``true" Chaplygin gas.

Plugging \eqref{eoszeroo} and \eqref{rhosolve} into \eqref{acceq} and \eqref{wq}, the deceleration parameter can be written as
\begin{equation}
q = \frac{1}{2}\left[1 - 3\,\frac{\mu}{\mu + \left(1 - \mu\right)a^{-3n/2}} \right]\;. \label{qmfin}
\end{equation}
Denoting the value of $a$ at $q=0$ by $a_{\rm q}$, we have 
\begin{equation}
\mu = \frac{1}{1 + 2 a_{\rm q}^{3n/2}} = \frac{1}{3}\left(1 - 2q_{0}\right)\ \Leftrightarrow\
q_0 = \frac{a_{\rm q} - 1}{2a_{\rm q} + 1}\;. \label{aqqo}
\end{equation}

In terms of either $a_{\rm q}$ or $q_0$ the pressure to energy density ratio takes the form
\begin{equation}
\frac{\Pi}{\rho} = - \frac{1}{1 + 2\left(\frac{a_{\rm q}}{a}\right)^{3n/2}} = - \frac{1 - 2q_{0}}{1 - 2q_{0} + 2\left(1+q_{0}\right)a^{-3n/2}}\;, \label{Prmu}
\end{equation}
and likewise the Hubble rate becomes
\begin{equation}
\frac{H}{H_{0}} = \left[\frac{1 + 2\left(\frac{a_{\rm q}}{a}\right)^{3n/2}}{1 + 2 a_{\rm q}^{3n/2}}\right]^{1/n} = \left(\frac{1}{3}\right)^{1/n}\left[1 - 2q_{0} + 2\left(1+q_{0}\right)a^{-3n/2}\right]^{1/n}\;. \label{Hqm}
\end{equation}

The parameters  $\zeta = \zeta(\rho)$ and $\tau = \tau(\rho)$ describe intrinsic properties of the fluid and their perturbations have the following forms:
\begin{equation}\label{zetataupertrho}
\delta\zeta = \frac{d\zeta(\rho)}{d\rho}\delta\rho\;, \qquad \delta\tau = \frac{d\tau(\rho)}{d\rho}\delta\rho\;.
\end{equation}
Equation~\eqref{eoszeroo} then provides us with the structure for $\zeta(\rho)$.
In the causal theories it will additionally allow us to calculate the relaxation time $\tau(\rho)$.

\subsection{Eckart's theory}

From \eqref{eoszeroo} and $\Pi = -\theta\zeta$ we can solve for $\zeta$ and find
\begin{equation}\label{zetaEckart}
 \zeta = \mu\rho_0^{n/2}\frac{\rho^{1 - n/2}}{3H} = \frac{\mu\rho_0^{n/2}}{\sqrt{24\pi G}}\;\rho^{\frac{1 - n}{2}}\;,
\end{equation}
where we have used Friedmann's equation \eqref{Feq} to express $H$ as function of $\rho$. Taking into account \eqref{zetataupertrho} we find then
\begin{equation}\label{zetaperteckart}
 \frac{\delta\zeta}{\zeta} = \frac{1 - n}{2}\frac{\delta\rho}{\rho}\;,
\end{equation}
which we are going to use in the perturbative analysis of the next section.

\subsection{Full M\"{u}ller-Israel-Stewart's theory}

In this section the $\bullet$ derivative reduces to the time derivative and will be denoted by a dot again.   Assuming $\cb^{2}$ to be constant, we find from \eqref{ghfunc}
\begin{equation}
 g = 2\;, \quad \mbox{ and } \quad h = \frac{1}{\rho}\;.
\end{equation}
With $\Pi/\rho$ given, equation \eqref{MISteviscousratio2} provides us with an expression for the parameter
$\tau H$:
\begin{equation}
3 \tau H = \frac{-\frac{\Pi}{\rho}}{\cb^2 + \frac{n}{2}\frac{\Pi}{\rho} + \frac{1}{2}\left(n-1\right)\left(\frac{\Pi}{\rho}\right)^{2}}\;.\label{tHfull}
\end{equation}
Using now \eqref{eoszeroo} and \eqref{Feq} we find for $\tau(\rho)$:
\begin{equation}
\tau(\rho) = \frac{1}{\sqrt{24\pi G\rho}}\;\frac{2\mu\left(\frac{\rho}{\rho_0}\right)^{-n/2}}{2\cb^2 - \mu n\left(\frac{\rho}{\rho_0}\right)^{-n/2} + \mu^2\left(n-1\right)\left(\frac{\rho}{\rho_0}\right)^{-n}}\;.\label{tHfullrho}
\end{equation}
Taking into account \eqref{zetataupertrho} we obtain
\begin{equation}\label{taupertfullMIS}
\frac{\delta\tau}{\tau} = -\frac{n}{2}\;\frac{\delta\rho}{\rho}\;\frac{2\cb^2 - \mu^2(n - 1)\left(\frac{\rho}{\rho_0}\right)^{-n}}{2\cb^2 - \mu n\left(\frac{\rho}{\rho_0}\right)^{-n/2} + \mu^2(n - 1)\left(\frac{\rho}{\rho_0}\right)^{-n}} - \frac{1}{2}\;\frac{\delta\rho}{\rho}\;.
\end{equation}
To make physical sense, the parameter $\tau H$ has to be positive. Combining \eqref{tHfull} and \eqref{Prmu},
the condition $\tau H > 0$ is equivalent to
\begin{equation}
\cb^{2} > \frac{1}{2}\frac{1 + 2n\left(\frac{a_{\rm q}}{a}\right)^{3n/2}}{\left[1 + 2\left(\frac{a_{\rm q}}{a}\right)^{3n/2}\right]^{2}} = \frac{1}{2}\left(1 - 2q_{0}\right)
\frac{1 - 2q_{0} + 2n\left(1 + q_{0}\right)a^{-3n/2}}
{\left[1 - 2q_{0} + 2\left(1 + q_{0}\right)a^{-3n/2}\right]^{2}}\;.
\label{cbcond}
\end{equation}
For arbitrary $a$, this is only guaranteed for $\cb^{2} > \frac{1}{2}$. Remarkably, this limit does not depend on $n$. As we shall see, such a large minimum value for the bulk viscous speed of sound has consequences for the perturbation dynamics, which disfavour the theory.

\subsection{Truncated M\"{u}ller-Israel-Stewart's theory}

In that case we have
\begin{equation}
\dot{\Pi} + \frac{1}{\tau}\Pi = -3H f\;, \label{dotPitr}
\end{equation}
instead of \eqref{MISteviscousratio2}. Solving for the relaxation time we obtain
\begin{equation}
3 \tau H = \frac{-\frac{\Pi}{\rho}}{\cb^2 - \left(1 - \frac{n}{2}\right)\frac{\Pi}{\rho}\left(1 + \frac{\Pi}{\rho}\right)}\;.\label{tHtrunc}
\end{equation}
Using again \eqref{eoszeroo} and \eqref{Feq} we find $\tau(\rho)$ in the truncated case:
\begin{equation}
\tau(\rho) = \frac{1}{\sqrt{24\pi G\rho}}\;\frac{2\mu\left(\frac{\rho}{\rho_0}\right)^{-n/2}}{2\cb^2 + \mu\left(2 - n\right)\left(\frac{\rho}{\rho_0}\right)^{-n/2}\left[1 - \mu\left(\frac{\rho}{\rho_0}\right)^{-n/2}\right]}\;.\label{tHtruncrho}
\end{equation}
With \eqref{zetataupertrho} it follows that
\begin{equation}\label{tauperttruncMIS}
\frac{\delta\tau}{\tau} = -\frac{n}{2}\;\frac{\delta\rho}{\rho}\;\frac{2\cb^2 - \mu^2(n - 2)\left(\frac{\rho}{\rho_0}\right)^{-n}}{2\cb^2 - \mu\left(n - 2\right)\left(\frac{\rho}{\rho_0}\right)^{-n/2} + \mu^2\left(n - 2\right)\left(\frac{\rho}{\rho_0}\right)^{-n}} - \frac{1}{2}\;\frac{\delta\rho}{\rho}\;.
\end{equation}
As far as the viscous speed of sound is concerned, the situation is different from that of the full theory.
Here we have
\begin{equation}
3\,H\,\tau = \frac{1 + 2\,\left(\frac{a_{\rm q}}{a}\right)^{3n/2}}{\cb^2\left[1 + 2\,\left(\frac{a_{\rm q}}{a}\right)^{3n/2}\right]^{2} + \left(2 - n\right)\,\left(\frac{a_{\rm q}}{a}\right)^{3n/2}}\;, \label{3Htau}
\end{equation}
or, equivalently,
\begin{equation}
3\,H\,\tau = \left(1 - 2q_{0}\right)
\frac{1 - 2q_{0} + 2\left(1 + q_{0}\right)a^{-3n/2}}
{\cb^{2}\left[1 - 2q_{0} + 2\left(1 + q_{0}\right)a^{-3n/2}\right]^{2}
+ \left(2 -
n\right)\left(1 - 2q_{0}\right)\,\left(1 + q_{0}\right)a^{-3n/2}}
\ . \label{3Htauq}
\end{equation}
In the truncated theory the requirement $3H\tau > 0$  implies that
\begin{equation}\label{cb2constrtrunc}
\cb^2 > \left(n - 2\right)\frac{\left(1 - 2q_{0}\right)\left(1 + q_{0}\right)a^{-3n/2}}
{1 - 2q_{0} + 2\left(1 + q_{0}\right)a^{-3n/2}}
\;,
\end{equation}
which (for $\cb^2 > 0$ ) is automatically satisfied for $n < 2$. In particular, at high redshifts, corresponding to $a \ll 1$, it follows that
\begin{equation}
3\,H\,\tau \approx
\frac{1}{2\,c_{b}^{2}}\,\frac{1 - 2q_{0}}{1 + q_{0}} a^{3n/2}
\qquad\qquad (a \ll 1)
 \;. \label{3Htaull}
\end{equation}
This means that either for $a \ll 1$ and $\cb^{2} \lesssim 1$ or for $\cb^{2} \ll 1$ with $\cb^{2} \gg a^{3n/2}$, one has
\begin{equation}
\tau \ll H^{-1}\;. \label{taull}
\end{equation}
A small relaxation time at high redshift is consistent with the limit $|\Pi|/\rho \ll 1$ for $a \ll 1$ [cf. \eqref{Prmu}]. 
In fact, during this period we have close-to-equilibrium conditions. 
If we assume our model to be valid from the epoch of matter-radiation equality
at $z_{\mathrm{eq}} \approx 3.3\cdot 10^{3}$ on, i.e. for $a \gtrsim a_{\mathrm{eq}} \approx 3\cdot 10^{-4}$, the condition for a model, similar to the $\Lambda$CDM model with $n\approx 2$, is $\cb^{2} \gg 10^{-11}$.
As we shall show below, acceptable values for $\cb^{2}$ are of the order of $\cb^{2} \lesssim  10^{-8}$, which is well compatible with that lower bound. This illustrates that a finite relaxation time requires a non-vanishing, albeit small, value of $\cb^{2}$. It is obvious from \eqref{3Htaull}, that $H\,\tau$ increases with the scale factor. 

At the present time $a = 1$ the expression (\ref{3Htauq}) specifies to
\begin{equation}
\tau H|_{0} =
\frac{1 - 2q_{0}}{9 \cb^{2} + \left(2 -
n\right)\left(1 - 2q_{0}\right)\,\left(1 + q_{0}\right)}
\ . \label{3Htau0q}
\end{equation}
We have
\begin{equation}
\tau|_{0} \gtrsim H^{-1}\;. \label{tau0}
\end{equation}
The relaxation time is of the order of the Hubble time or even exceeds the latter. Finally, in the long time limit, i.e., $a \gg 1$, relation \eqref{3Htauq} becomes
\begin{equation}
\tau \approx \frac{1}{3\,\cb^{2}}\, H^{-1} \qquad\qquad (a \gg 1)\;. \label{taug}
\end{equation}
It follows, that for any $\cb^{2} < 1/3$, but larger than a very small minimum value $\approx 10^{-10}$, the relaxation time
will remain finite and larger than the Hubble time, i.e., $\tau \gtrsim H^{-1}$ or even $\tau \gg H^{-1}$.

The situation is different for $n > 2$. Equation~\eqref{cb2constrtrunc} is equivalent to the existence of a lower limit on $\cb^2$. More precisely, the maximum of the function on the right hand side of \eqref{cb2constrtrunc} occurs at
\begin{equation}
 \left(\frac{a_{\rm q}}{a}\right)^{3n/2} = \frac{1}{2}\;.
\end{equation}
Therefore, the relevant limit is
\begin{equation}
\cb^2 > \frac{n - 2}{8}\;.
\end{equation}
For example, for the Chaplygin gas ($n = 4$), it amounts to $\cb^{2} > 1/4$. This value is slightly smaller than that of the full theory but, as the numerical analysis will show, unless $n$ is extremely close to $2$, this limit will not admit a favourable scenario.

In the next section we shall see that a perturbation theory based on the truncated version for $n\leq 2$ fares better with respect to the behaviour of the gravitational potential than the full theory. Different from the full theory, the truncated theory allows for a finite but small value $\cb^{2} \ll 1$.

It is expedient to recall that both the full and the truncated theories are applied here under conditions where a microscopic justification is missing. Both versions were taken here on a purely phenomenological basis. Intuitively, the simpler truncated version appears even more transparent since it represents the minimally possible generalisation of Eckart's theory. In the present context, the truncated theory is not necessarily an approximation to the full theory, but it should be considered as a phenomenological description on its own. From this point of view one should not be too surprised to obtain ``better" results from the truncated theory.


\section{Results: the evolution of the gravitational potential}\label{Sec:Results}

In this section we numerically solve the equation for the gravitational potential, for the Eckart, the MIS and the truncated MIS cases. Independently from the transport theory, the background evolution is the same, viz. \eqref{Hqm}, due to our choice \eqref{eoszeroo}. In particular, for the  case corresponding to the $\Lambda$CDM model,  $n = 2$, we have
\begin{equation}
\frac{H^2}{H^2_{0}} = \frac{1 + 2\left(\frac{a_{\rm q}}{a}\right)^{3}}{1 + 2 a_{\rm q}^{3}}\;,
\end{equation}
 with the $\Lambda$CDM identifications
\begin{equation}
 \Omega_{\Lambda 0} = \frac{1}{1 + 2 a_{\rm q}^{3}}\;, \qquad \Omega_{\rm m0} = \frac{2a_{\rm q}^3}{1 + 2 a_{\rm q}^{3}} = 1 - \Omega_{\Lambda 0}\;.
\end{equation}
For $\Omega_{\Lambda 0} = 0.734$, we have $a_{\rm q} = 0.566$, or a redshift $z_{\rm q} = 0.767$ and $q_0 = -0.204$.
In this section we denote the derivative wrt the scale factor by a subscript $a$.

\subsection{Eckart's case}

Linking $\delta\rho$ to the gravitational potential via the $(0-0)$ Einstein equation \eqref{00EE}, we get from \eqref{zetaperteckart}
\begin{equation}\label{deltaxiEckart}
 \frac{\delta\zeta}{\zeta} = \frac{1 - n}{3\h^2}\left[\Delta\Phi - 3\h\left(\Phi' + \h\Phi\right)\right]\;.
\end{equation}
Using now the $(0-i)$ Einstein equation \eqref{0iEE} written as
\begin{equation}\label{ThetaEckart}
\Theta = \frac{\Delta\left(\Phi' + \h\Phi\right)}{\h' - \h^2}\;,
\end{equation}
together with \eqref{dPiEckart}, \eqref{deltaxiEckart} and the $(i-j)$ Einstein equation \eqref{ijEE}, we obtain
\begin{equation}
 \Phi'' + 3\h\Phi' + \left(\h^2 + 2\h'\right)\Phi = 4\pi G a \zeta\left[\frac{n - 1}{\h}\Delta\Phi - \frac{\Delta\left(\Phi' + \h\Phi\right)}{\h' - \h^2} - 3(n - 2)\left(\Phi' + \h\Phi\right)\right]\;.
\end{equation}
 In terms of derivatives wrt the scale factor (subscript $a$)  we get
\begin{eqnarray}\label{GravPotEckeq0}
 \Phi_{aa} + \left(\frac{4}{a} + \frac{\h_a}{\h}\right)\Phi_a + \left(\frac{1}{a^2} + \frac{2\h_a}{a\h}\right)\Phi = \nonumber\\ \frac{4\pi G\zeta}{a\h^3}\left[(n - 1)\Delta\Phi - \frac{\h\Delta\left(a\Phi_a + \Phi\right)}{a\h_a - \h} - 3\h^2(n - 2)\left(a\Phi_a + \Phi\right)\right]\;.
\end{eqnarray}
Performing a plane wave expansion $\Phi \propto \exp\left(i{\bf k\cdot x}\right)$ and using \eqref{zetaEckart}, we can cast \eqref{GravPotEckeq0} in the form 
\begin{eqnarray}\label{GravPotEckeq}
\Phi_{aa} + \left(\frac{4}{a} + \frac{\h_a}{\h}\right)\Phi_a + \left(\frac{1}{a^2} + \frac{2\h_a}{a\h}\right)\Phi = \nonumber\\ -\frac{\left(1 - 2q_0\right)H_0^nk^2}{6\h^{n + 2} a^{2 - n}}\left[(n - 1)\Phi - \frac{\h\left(a\Phi_a + \Phi\right)}{a\h_a - \h}\right] - \frac{\left(1 - 2q_0\right)H_0^n}{2\h^{n} a^{2 - n}}(n - 2)\left(a\Phi_a + \Phi\right)\;.
\end{eqnarray}
For the  case that simulates the $\Lambda$CDM model, $n = 2$, Eq.~\eqref{GravPotEckeq} becomes
\begin{equation}\label{GravPotEckeqLCDM}
\Phi_{aa} + \left(\frac{4}{a} + \frac{\h_a}{\h}\right)\Phi_a + \left(\frac{1}{a^2} + \frac{2\h_a}{a\h}\right)\Phi = -\frac{\left(1 - 2q_0\right)H_0^2k^2}{6\h^4}\left(\Phi - \h\frac{a\Phi_a + \Phi}{a\h_a - \h}\right)\;.
\end{equation}
The $\Lambda$CDM model is recovered for a vanishing right-hand side. A non-vanishing right-hand side
is the result of pressure perturbations. It is a characteristic feature of the viscous model, that the $k^{2}$ term in \eqreff{GravPotEckeq} does not only contain $\Phi$ but the derivative $\Phi_{a}$ as well.
This coupling to $\Phi_{a}$ (or its energy-density counterpart) turned out to be essential for an adequate description of the matter power spectrum \cite{HipolitoRicaldi:2009je,HipolitoRicaldi:2010mf}.
Generally speaking, for the behaviour of the gravitational potential such a coupling causes deviations
from  the $\Lambda$CDM model which are especially relevant  from the viewpoint of the integrated Sachs-Wolfe (ISW) effect \cite{Sachs:1967er}, where an integration of $\dot{\Phi}$ up to $k \to \infty$ has to be performed.

\begin{figure}[t]
\begin{center}
\includegraphics[width=0.5\columnwidth]{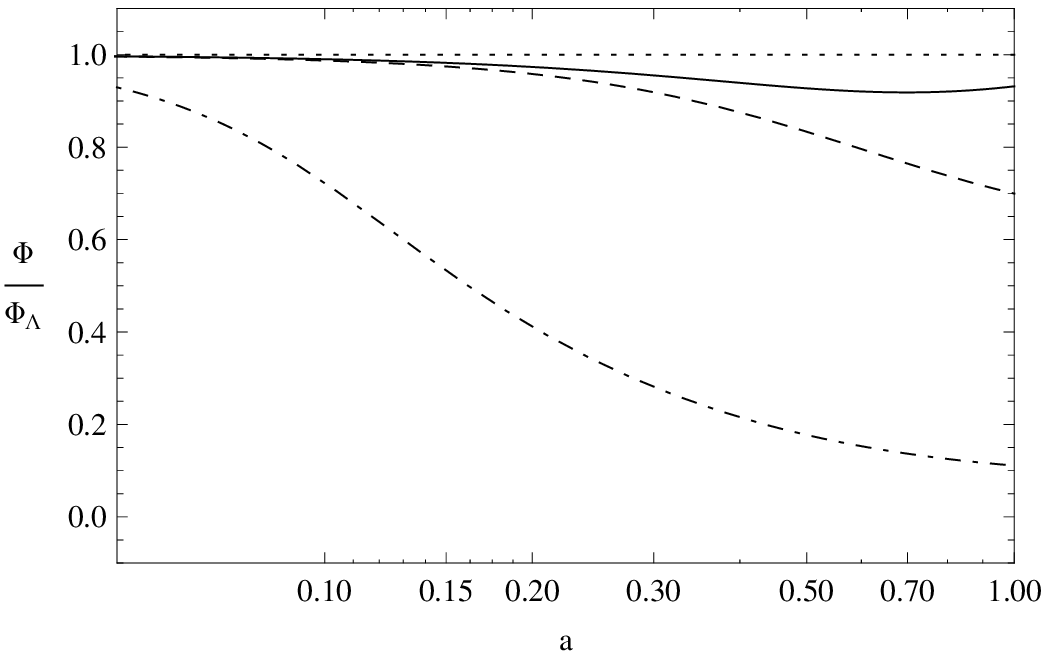}\includegraphics[width=0.5\columnwidth]{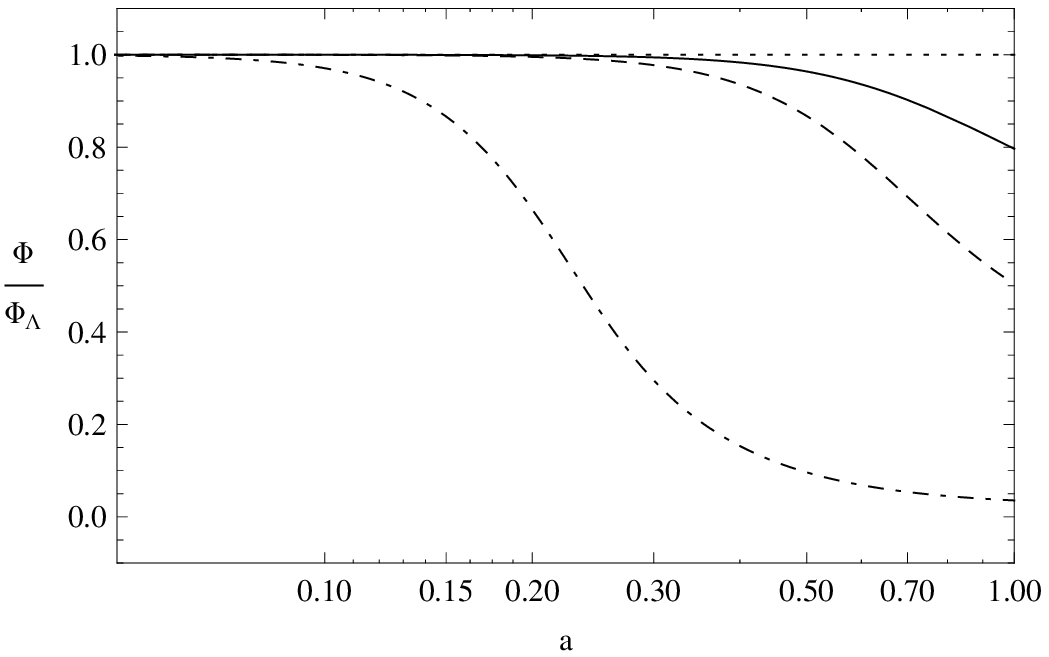}\\
\includegraphics[width=0.5\columnwidth]{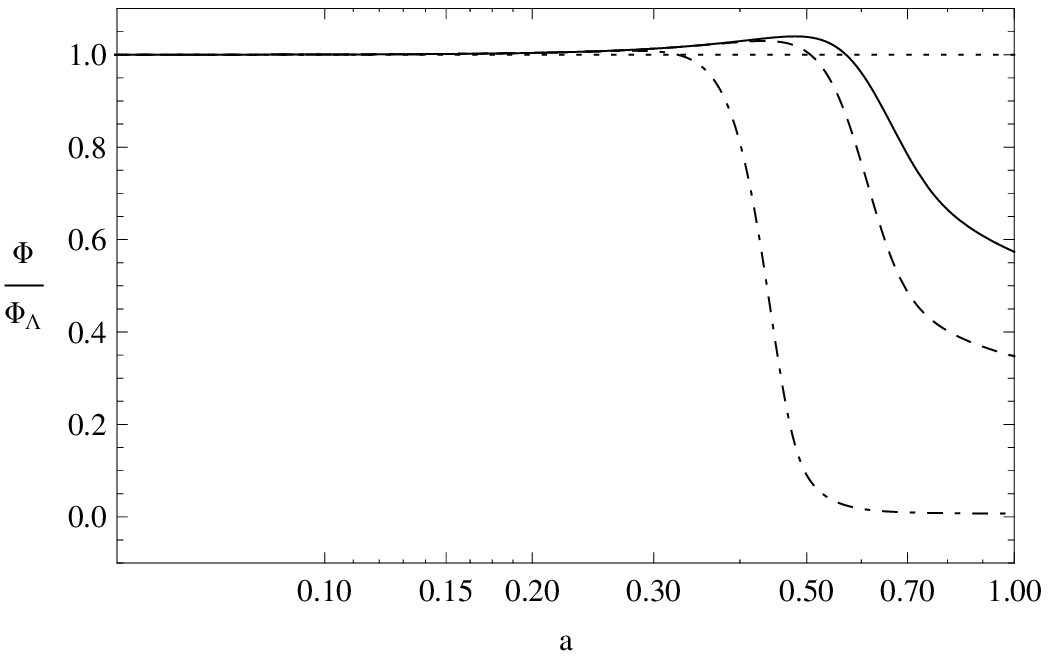}\includegraphics[width=0.5\columnwidth]{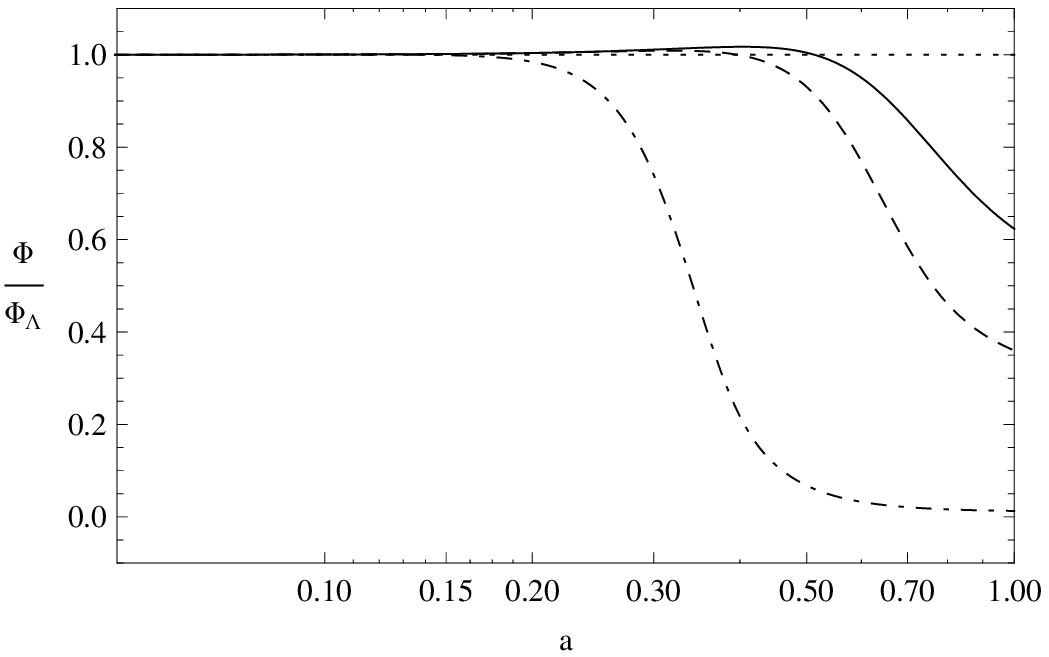}
\caption{Ratio $\Phi/\Phi_\Lambda$ in Eckart's theory as function of the scale factor for $k = 5\cdot 10^{-4}, 10^{-3}, 10^{-2}$ $h$ Mpc$^{-1}$ (solid, dashed, dot-dashed lines, respectively). From the upper-left panel in clockwise sense, $n = 1, 2, 4, 8$.}
\label{fig1}
\end{center}
\end{figure}
\begin{figure}[t]
\begin{center}
\includegraphics[width=0.5\columnwidth]{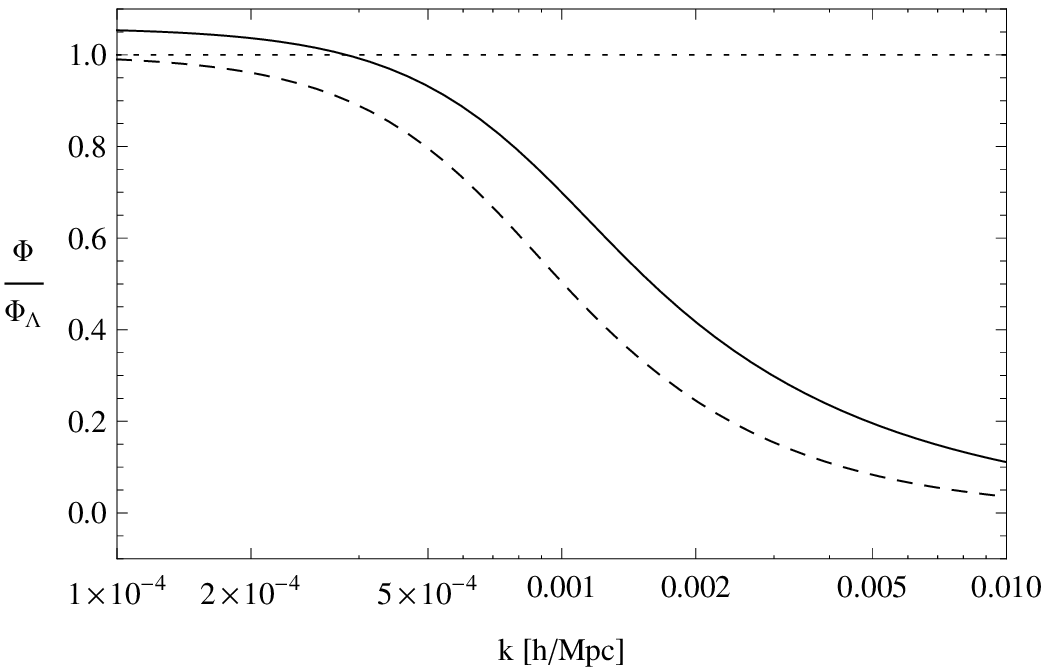}\includegraphics[width=0.5\columnwidth]{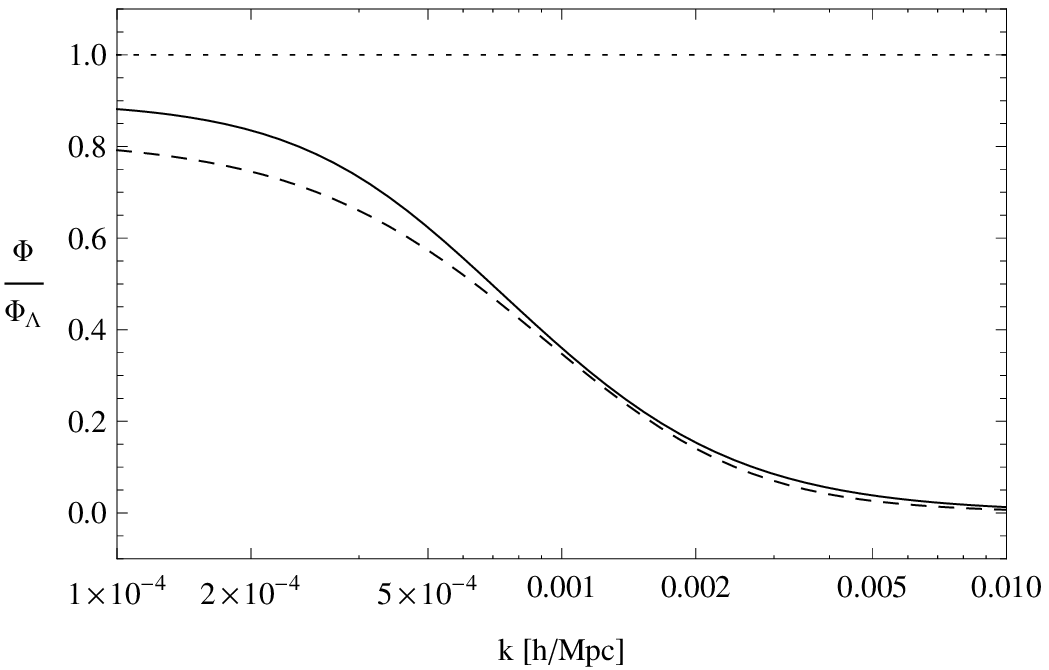}\\
\caption{Ratio $\Phi/\Phi_\Lambda$ in Eckart's theory as function of the wavenumber $k$ for $a = 1$ fixed. {\it Left Panel:} $n = 1, 2$ (solid, dashed). {\it Right panel:} $n = 4, 8$ (solid, dashed).}
\label{fig1bis}
\end{center}
\end{figure}

Numerical solutions of \eqreff{GravPotEckeq} are displayed in Fig.~\ref{fig1} and Fig.~\ref{fig1bis}. We plot the ratio of the viscous gravitational potential computed from \eqreff{GravPotEckeq} to the $\Lambda$CDM one (which is also computed from \eqreff{GravPotEckeq} with $k = 0$ and $n = 2$). We have chosen initial conditions $\Phi(a_*) = 1$, $\dot{\Phi}(a_*) = 0$, where $a_* = 10^{-3}$ is the recombination scale factor. Moreover, we have also chosen $a_{\rm q} = 0.566$, as the reference transition redshift of the $\Lambda$CDM model.

From the plot in Figs.~\ref{fig1} and \ref{fig1bis} we can infer the following features:
\begin{enumerate}
\item  On the largest scales and for small values of the scale factor, the results of Eckart's theory approach those of the $\Lambda$CDM model. On smaller scales and for scale factors of the order of one, the differences are substantial or even dramatic. Given, that the equations just differ by $k^{2}$ terms, this is not surprising. 
\item Varying the background expansion implies the following: for increasing $n$, the effect of the viscosity contribution  on intermediate and smaller scales seems to be ``delayed'' and the agreement  with the $\Lambda$CDM result enhances for $a<a_{\rm q}$.  On the other hand, for $a>a_{\rm q}$ the differences increase.
\item The shape of $\Phi/\Phi_\Lambda$  close to $a = 1$ in Fig.~\ref{fig1bis}  does not depend much on $n$. An increasing $n$  causes only a shift towards smaller values of the ratio $\Phi/\Phi_\Lambda$.
\end{enumerate}
Already from this qualitative analysis we may expect the agreement of this model with observation to be poor.

\newpage

\subsection{Full M\"{u}ller-Israel-Stewart's theory}

With our ansatz $f = \cb^2\rho$ and $\cb^2$ constant (which implies $g = 2$ and $h = 1/\rho$), \eqreff{MISteviscousratio3pert} can be cast into the following form:
\begin{eqnarray}\label{PipertvofullMIS}
\left(\frac{\delta\Pi}{\rho}\right)_a = - \frac{1}{Ha\tau}\frac{\delta\Pi}{\rho} + \frac{1}{Ha\tau}\frac{\Pi}{\rho}\frac{\delta\tau}{\tau} -\frac{3}{a}\Phi\left(1 - \frac{n}{2}\right)\frac{\Pi}{\rho}\left(1 + \frac{\Pi}{\rho}\right)\nonumber\\
- \frac{\delta\theta}{Ha}\left(\cb^2 + \frac{\Pi}{\rho} + \frac{\Pi^2}{2\rho^2}\right) - \frac{3}{a}\frac{\delta\rho}{\rho}\left(\cb^2 - \frac{\Pi^2}{2\rho^2}\right)\;,
\end{eqnarray}
where the subscript $a$ again denotes derivation wrt the scale factor. The evolution of $\Pi/\rho$ in \eqref{PipertvofullMIS} is known by \eqref{Prmu}. Taking into account \eqref{taupertfullMIS} we find the following system of coupled equations:
\begin{eqnarray}
\label{dPhifullMIS}\Phi_{aa} &+& \left(\frac{4}{a} + \frac{\h_a}{\h}\right)\Phi_a + \left(\frac{1}{a^2} + \frac{2\h_a}{a\h}\right)\Phi = \frac{3}{2a^2}\left(\frac{\delta\Pi}{\rho}\right)\;,\\
\label{dPIoAfullMIS}\left(\frac{\delta\Pi}{\rho}\right)_a &=& \frac{3}{a}\left(2 - n\right)\Phi\left[\cb^2 - \frac{(n + 1)\Pi^2}{2\rho^2}\right] + \frac{3}{a}\left[\cb^2 + \frac{n\Pi}{2\rho} + \frac{(n - 1)\Pi^2}{2\rho^2}\right]\frac{\delta\Pi}{\Pi}\nonumber\\ &+& 3(2 - n)\Phi_a\left(\cb^2 + \frac{\Pi}{2\rho} - \frac{n\Pi^2}{2\rho^2}\right) + \frac{(2 - n)k^2}{a\h^2}\Phi\left(\cb^2 + \frac{\Pi}{2\rho} - \frac{n\Pi^2}{2\rho^2}\right) \nonumber\\ &-& \left(\cb^2 + \frac{\Pi}{\rho} + \frac{\Pi^2}{2\rho^2}\right)\frac{k^2}{a\h^2}\left(\Phi - \h\frac{a\Phi_a + \Phi}{a\h_a - \h}\right)\;.
\end{eqnarray}
Obviously, the simplest case is  $n = 2$, the case which reproduces the $\Lambda$CDM background. It eliminates various contributions,  among them also a $k^2\Phi$ term. However, the above system of equations is so complicated that only a numerical analysis will uncover its secrets. We have now one parameter more than in Eckart's theory, the constant $\cb^2$. We present numerical results of the above system, recalling the limit $\cb^2 > 1/2$ [see the comment following \eqref{cbcond}].
As for Eckart's case, we choose $\Phi(a_*) = 1$, $\dot{\Phi}(a_*) = 0$ as initial conditions on the gravitational potential plus $\delta\Pi(a_*) = 0$, i.e. assuming that perturbations in the viscous pressure of the fluid were not important at recombination. Again, we choose $a_{\rm q} = 0.566$ as the transition redshift. We plot the results for different choices of the background evolution, i.e. for various $n$'s, and for different choices of $\cb^2 > 1/2$.

\begin{figure}[t]
\begin{center}
\includegraphics[width=0.5\columnwidth]{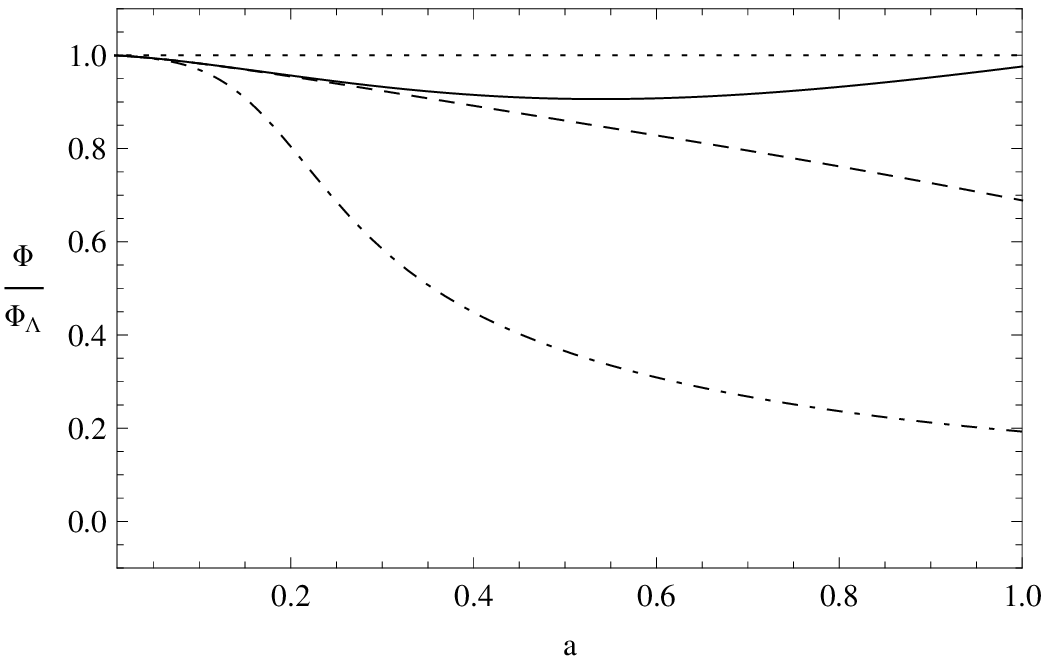}\includegraphics[width=0.5\columnwidth]{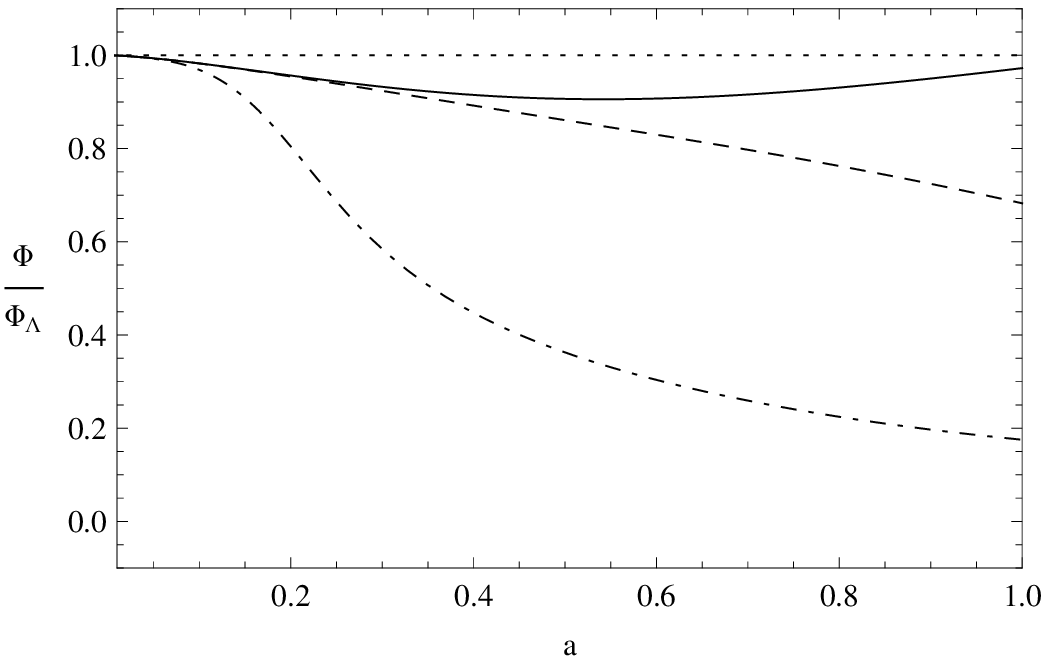}\\
\includegraphics[width=0.5\columnwidth]{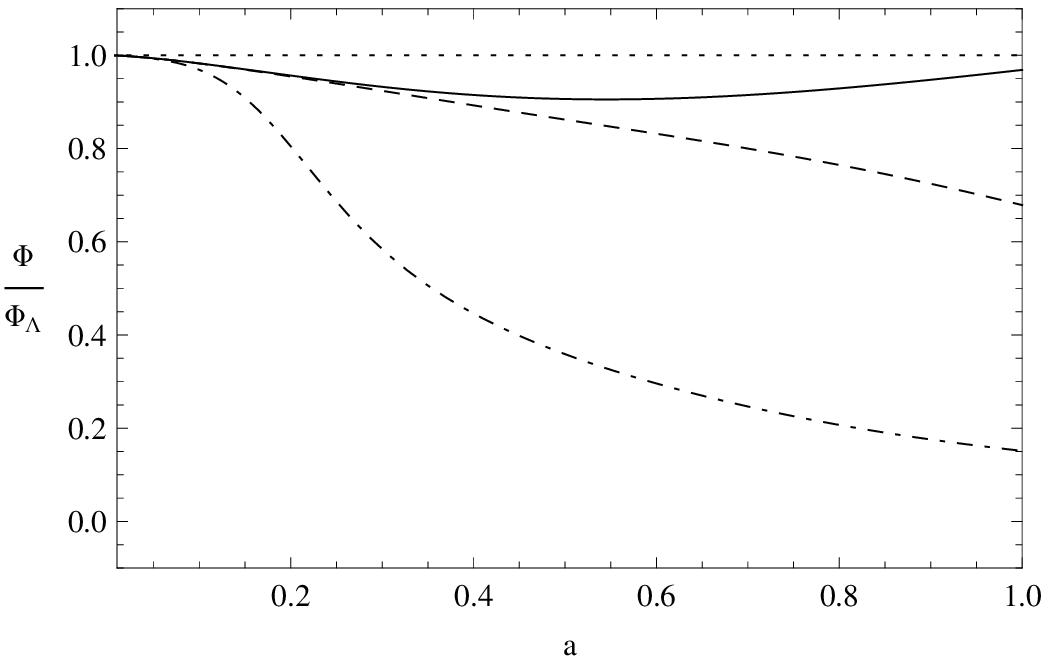}\includegraphics[width=0.5\columnwidth]{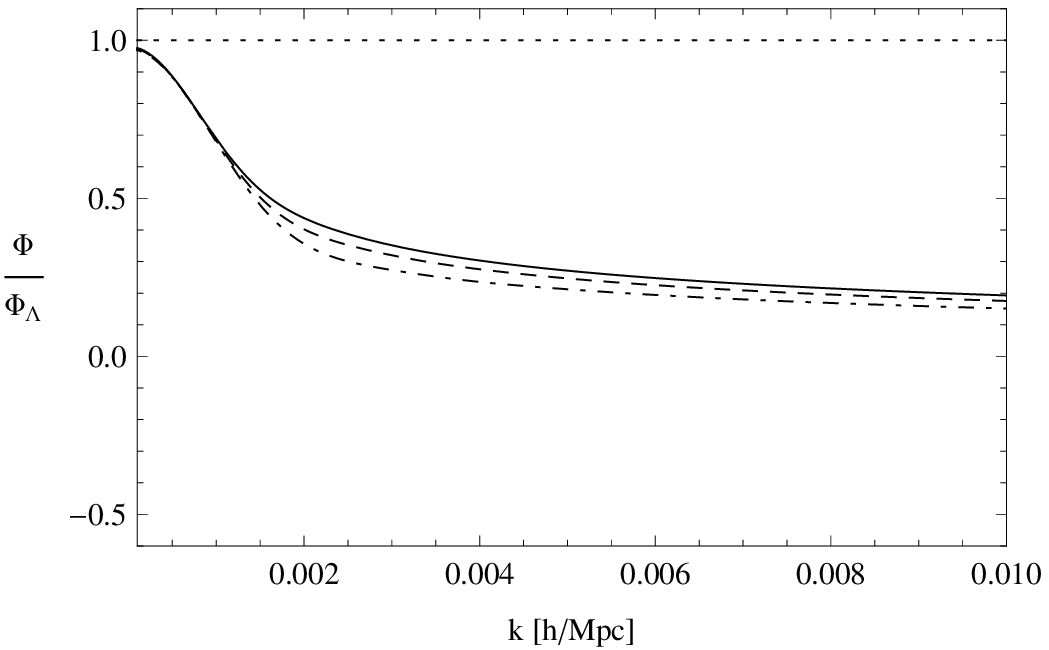}
\caption{{\bf Results for the case ${\bf n = 1}$:} ratio $\Phi/\Phi_\Lambda$ in the full MIS theory for $f(\rho) = \cb^2\rho$ with $\cb^2$ constant. {\it Top-left panel:} $\cb^2 = 1$. {\it Top-right panel:} $\cb^2 = 0.7$. {\it Bottom-left panel:} $\cb^2 = 0.5$. In the latter three panels the profiles are functions of $a$ for $k = 0.0001, 0.001, 0.01$ $h$ Mpc$^{-1}$ fixed (solid, dashed, dot-dashed, respectively). {\it Bottom-right panel:}  dependence on  $k$ for $a = 1$; here the solid, dashed, dot-dashed lines correspond to the cases $\cb^2 = 1, 0.7, 0.5$, respectively.}
\label{fig3}
\end{center}
\end{figure}

\begin{figure}[t]
\begin{center}
\includegraphics[width=0.5\columnwidth]{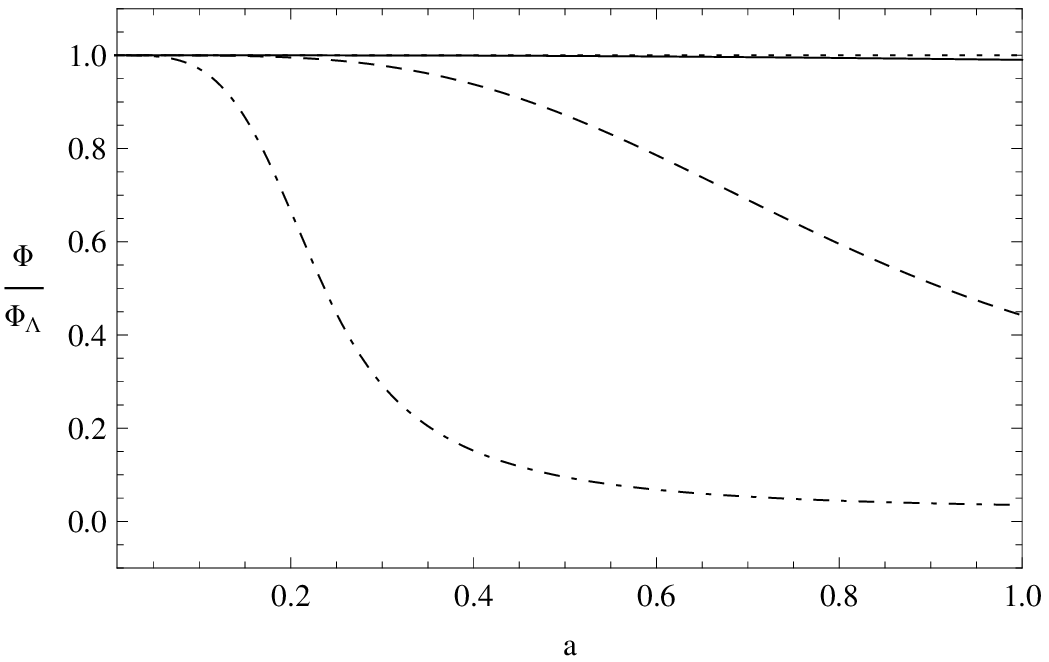}\includegraphics[width=0.5\columnwidth]{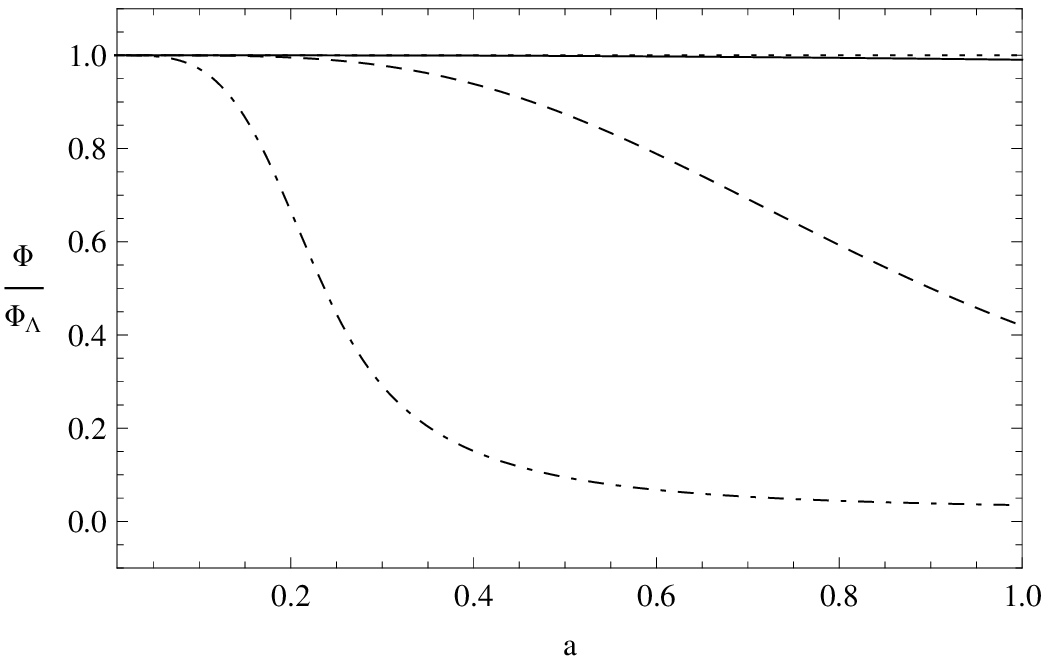}\\
\includegraphics[width=0.5\columnwidth]{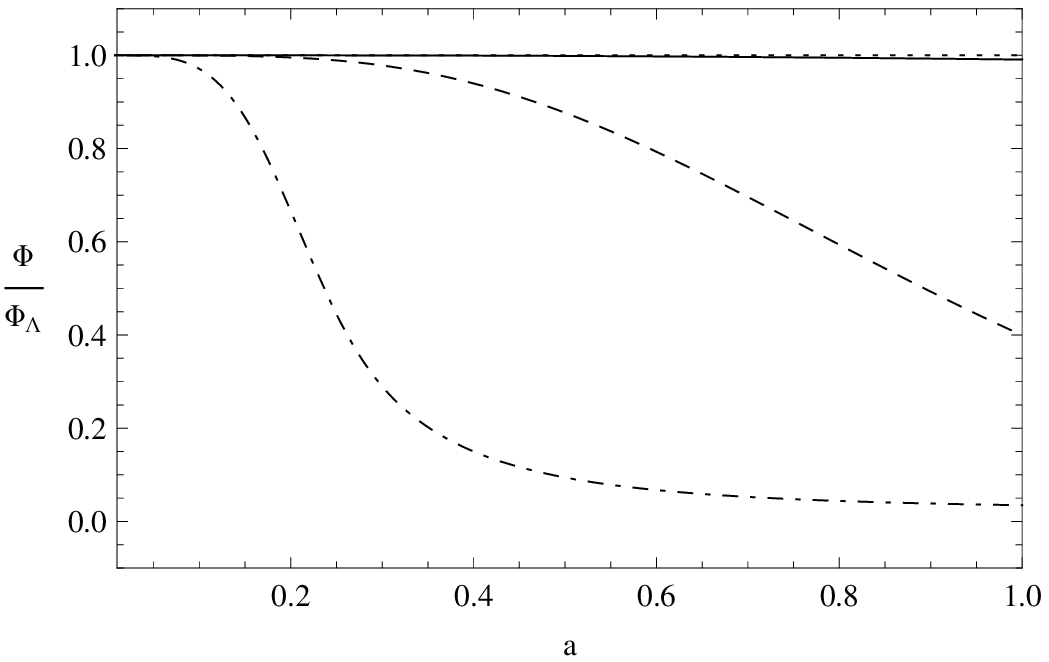}\includegraphics[width=0.5\columnwidth]{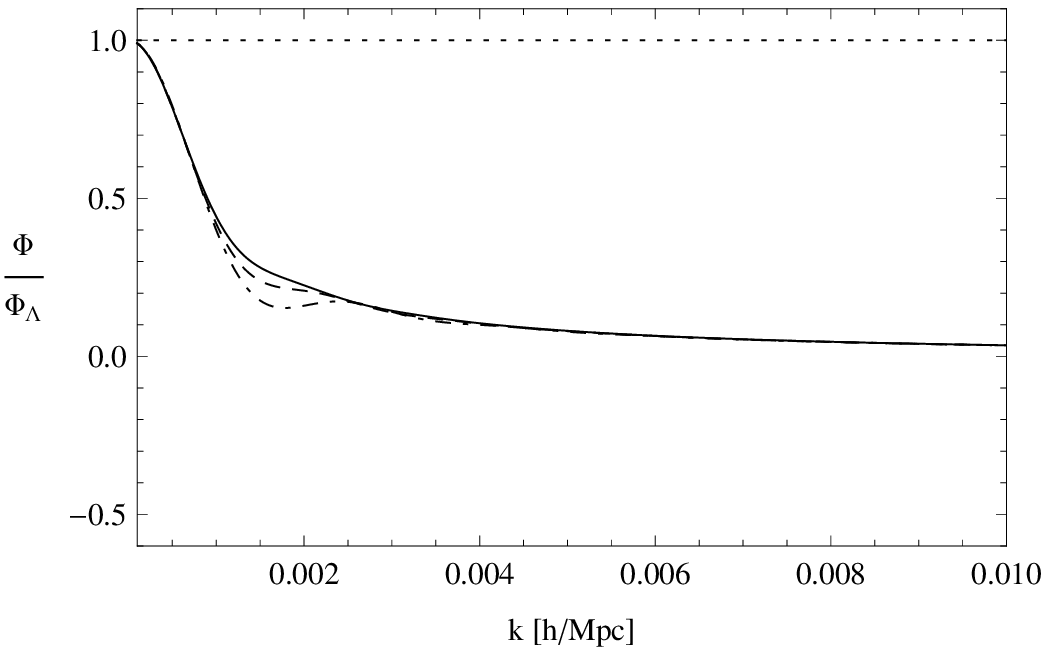}
\caption{{\bf Results for the case ${\bf n = 2}$, i.e. a $\Lambda$CDM background:} ratio $\Phi/\Phi_\Lambda$ in the full MIS theory for $f(\rho) = \cb^2\rho$ with $\cb^2$ constant. {\it Top-left panel:} $\cb^2 = 1$. {\it Top-right panel:} $\cb^2 = 0.7$. {\it Bottom-left panel:} $\cb^2 = 0.5$. In the latter three panels the profiles are functions of $a$ for $k = 0.0001, 0.001, 0.01$ $h$ Mpc$^{-1}$ fixed (solid, dashed, dot-dashed, respectively). {\it Bottom-right panel:}  dependence on  $k$ for $a = 1$; here the solid, dashed, dot-dashed lines correspond to the cases $\cb^2 = 1, 0.7, 0.5$, respectively.}
\label{fig3bis}
\end{center}
\end{figure}

\begin{figure}[t]
\begin{center}
\includegraphics[width=0.5\columnwidth]{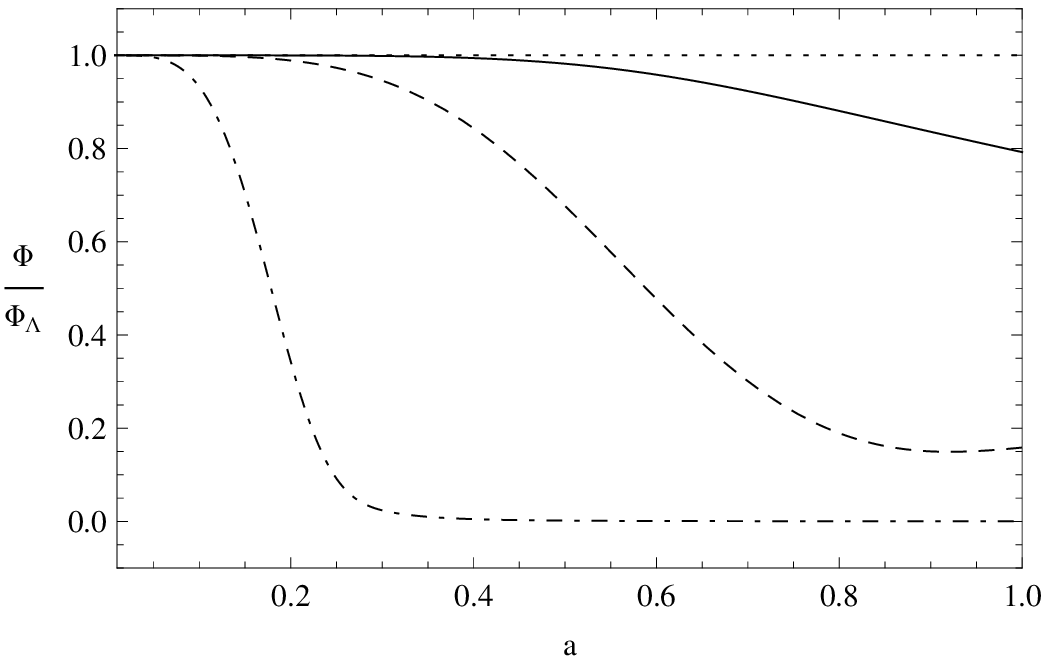}\includegraphics[width=0.5\columnwidth]{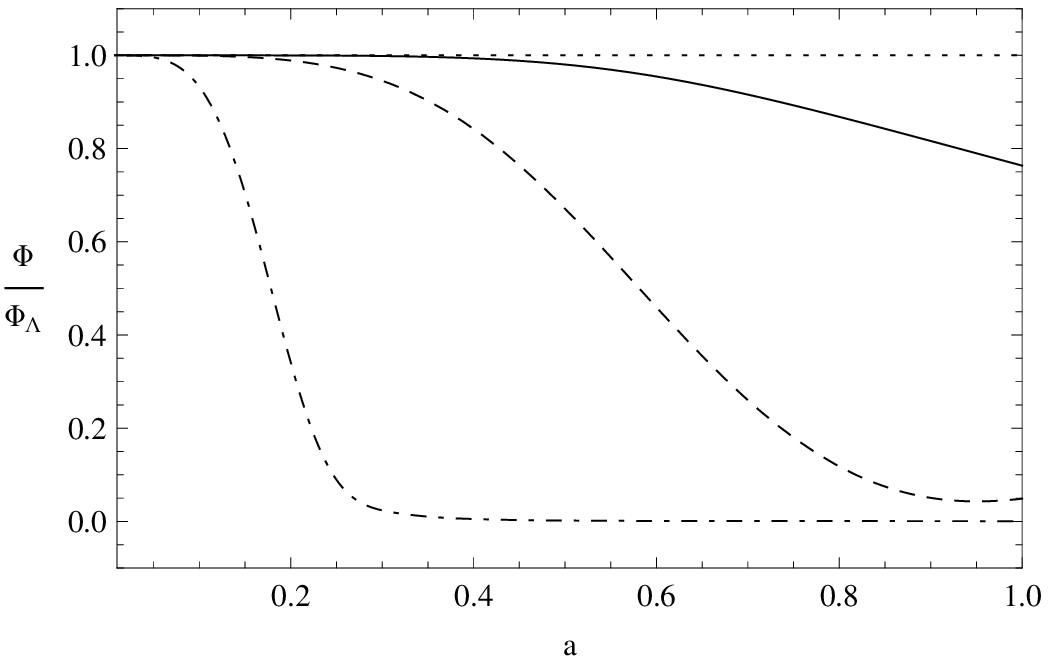}\\
\includegraphics[width=0.5\columnwidth]{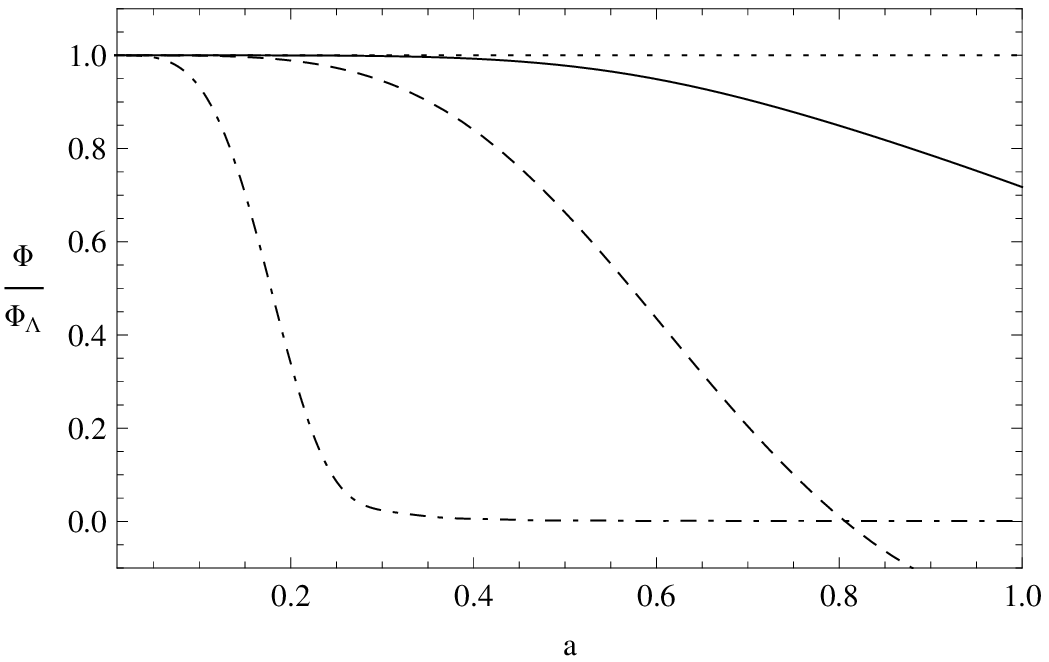}\includegraphics[width=0.5\columnwidth]{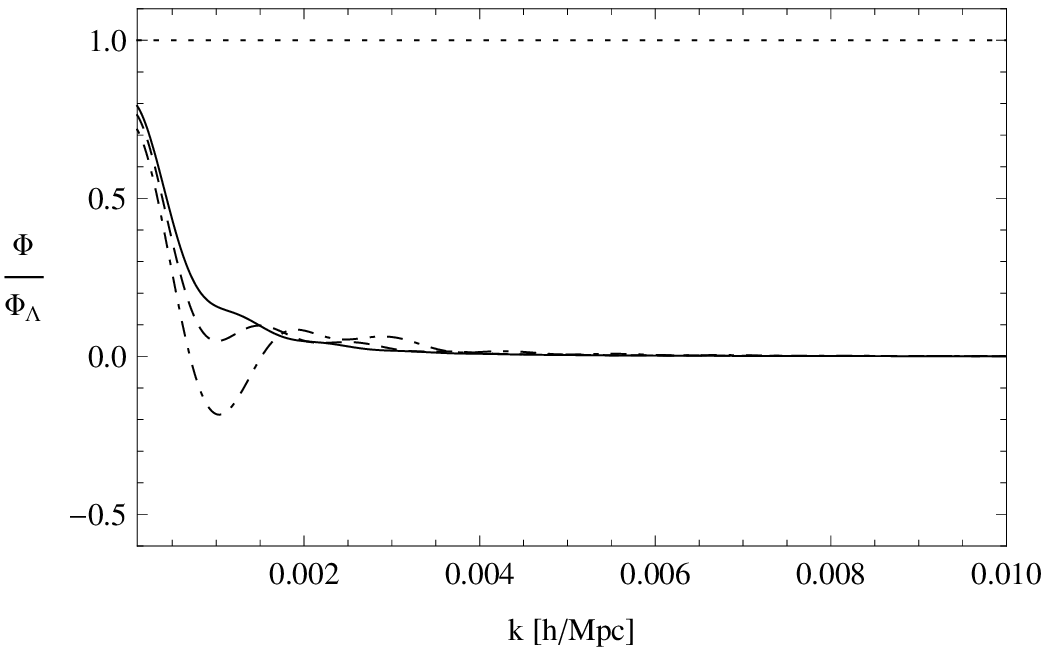}
\caption{{\bf Results for the case ${\bf n = 4}$:} ratio $\Phi/\Phi_\Lambda$ in the full MIS theory for $f(\rho) = \cb^2\rho$ with $\cb^2$ constant. {\it Top-left panel:} $\cb^2 = 1$. {\it Top-right panel:} $\cb^2 = 0.7$. {\it Bottom-left panel:} $\cb^2 = 0.5$. In the latter three panels the profiles are functions of $a$ for $k = 0.0001, 0.001, 0.01$ $h$ Mpc$^{-1}$ fixed (solid, dashed, dot-dashed, respectively). {\it Bottom-right panel:}   dependence on  $k$ for $a = 1$; here the solid, dashed, dot-dashed lines correspond to the cases $\cb^2 = 1, 0.7, 0.5$, respectively.}
\label{fig3ter}
\end{center}
\end{figure}
We infer the following from Figures~\ref{fig3}, \ref{fig3bis} and \ref{fig3ter}:
\begin{enumerate}
\item As for Eckart's case, the ratio $\Phi/\Phi_\Lambda$ decays faster and earlier in time for larger values of $k$.
\item For fixed $k$ and increasing $n$ the decay phase of the evolution is more pronounced and takes place earlier than in Eckart's case.
\item  Apparently, the value of $\cb^2$ has a very little influence on the evolution of the gravitational potential. This is probably due to the fact that $\cb^2$ must be larger than one-half, therefore already too large for allowing for a perturbative evolution close to the $\Lambda$CDM case. For $n=4$, a decreasing $\cb^2$ even enlarges the difference between  $\Phi$ and $\Phi_\Lambda$.
\end{enumerate}
In conclusion, it seems that taking into account the full causal theory does not amend the problems already encountered in Eckart's case.

\newpage

\subsection{Truncated M\"{u}ller-Israel-Stewart's theory}

With our ansatz $f = \cb^2\rho$ and $\cb^2$ constant, \eqreff{MISteviscousratio2perttrunc} can be cast into the following form:
\begin{eqnarray}
\left(\frac{\delta\Pi}{\rho}\right)_a - \frac{3\cb^2}{a}\frac{\delta\Pi}{\Pi} - \frac{3}{a}\left(1 + \frac{\Pi}{\rho}\right)\frac{n}{2}\frac{\delta\Pi}{\rho} + \frac{3}{a}\left[\cb^2 - \frac{\Pi}{\rho}\left(1 - \frac{n}{2}\right)\left(1 + \frac{\Pi}{\rho}\right)\right]\frac{\delta\tau}{\tau} = \nonumber\\ -\frac{3}{a}\cb^2\left(\frac{\delta\rho}{\rho} + \frac{\delta\theta}{\theta}\right) - \frac{3}{a}\Phi\frac{\Pi}{\rho}\left(1 - \frac{n}{2}\right)\left(1 + \frac{\Pi}{\rho}\right)\;,
\end{eqnarray}
where a subscript $a$ denotes a derivative wrt the scale factor. Together with \eqref{tauperttruncMIS}, we find the following system of coupled equations:
\begin{eqnarray}
\label{dPhitruncMIS}\Phi_{aa} &+& \left(\frac{4}{a} + \frac{\h_a}{\h}\right)\Phi_a + \left(\frac{1}{a^2} + \frac{2\h_a}{a\h}\right)\Phi = \frac{3}{2a^2}\left(\frac{\delta\Pi}{\rho}\right)\;,\\
\label{dPIoAtruncMIS}\left(\frac{\delta\Pi}{\rho}\right)_a &=& \frac{3\left(2 - n\right)}{a}\Phi\left[\cb^2 - \frac{n\Pi^2}{2\rho^2}\right] + \frac{3}{a}\left[\cb^2 + \frac{n\Pi}{2\rho}\left(1 + \frac{\Pi}{\rho}\right)\right]\frac{\delta\Pi}{\Pi}\nonumber\\ &+& 3(2 - n)\Phi_a\left[\cb^2 + \frac{\Pi}{2\rho} - (n - 1)\frac{\Pi^2}{2\rho^2}\right] + \frac{(2 - n)k^2}{a\h^2}\Phi\left[\cb^2 + \frac{\Pi}{2\rho} - (n - 1)\frac{\Pi^2}{2\rho^2}\right] \nonumber\\ &-& \frac{\cb^2k^2}{a\h^2}\left(\Phi - \h\frac{a\Phi_a + \Phi}{a\h_a - \h}\right)\;.
\end{eqnarray}
Also here, the case $n = 2$, corresponding to a $\Lambda$CDM background, turns out to be the simplest. Introducing $\Pi = -\mu\rho_0$ constant and the Friedmann equation, we obtain
\begin{eqnarray}
\label{dPhitruncMISn2}\Phi_{aa} &+& \left(\frac{4}{a} + \frac{\h_a}{\h}\right)\Phi_a + \left(\frac{1}{a^2} + \frac{2\h_a}{a\h}\right)\Phi = \frac{3}{2a^2}\left(\frac{\delta\Pi}{\rho}\right)\;,\\
\label{dPIoAtruncMISn2}\left(\frac{\delta\Pi}{\rho}\right)_a &=& - \frac{3}{a}\left[\frac{1 - \mu}{\mu a^{3} + (1 - \mu)}\right]\frac{\delta\Pi}{\rho} \nonumber\\ &-& \frac{3}{a}\cb^2\left[\frac{\mu + (1 - \mu)a^{-3}}{\mu}\frac{\delta\Pi}{\rho} + \frac{k^2\mu}{3\h^2}\left(\Phi - \h\frac{a\Phi_a + \Phi}{a\h_a - \h}\right)\right]\;,
\end{eqnarray}
where we have  separated the $\cb^2$ terms. In the limit $\cb^2 =0$ the right-hand side 
of \eqref{dPhitruncMISn2} rapidly decays and we recover the $\Lambda$CDM dynamics. However, $\cb^2 =0$ is equivalent to $f = \zeta = 0$, i.e., such a case does no longer describe a bulk viscous fluid. This corresponds to the circumstance that there also exists a lower limit for $\cb^2$ [see below \eqref{taull}], derived under conditions where the credibility of the theory is highest. Consequently, what is needed, is a very small but non vanishing value of $\cb^2$.\\
Let us consider how small $\cb^2$ has to be in order to have a viable truncated viscous theory. We choose $\Phi(a_*) = 1$, $\dot{\Phi}(a_*) = 0$ and $\delta\Pi(a_*) = 0$ as initial conditions, and $a_{\rm q} = 0.566$. We plot the results for different choices of the background evolution, i.e. for various $n$'s, and for different choices of $\cb^2$.\\
\begin{figure}[t]
\begin{center}
\includegraphics[width=0.5\columnwidth]{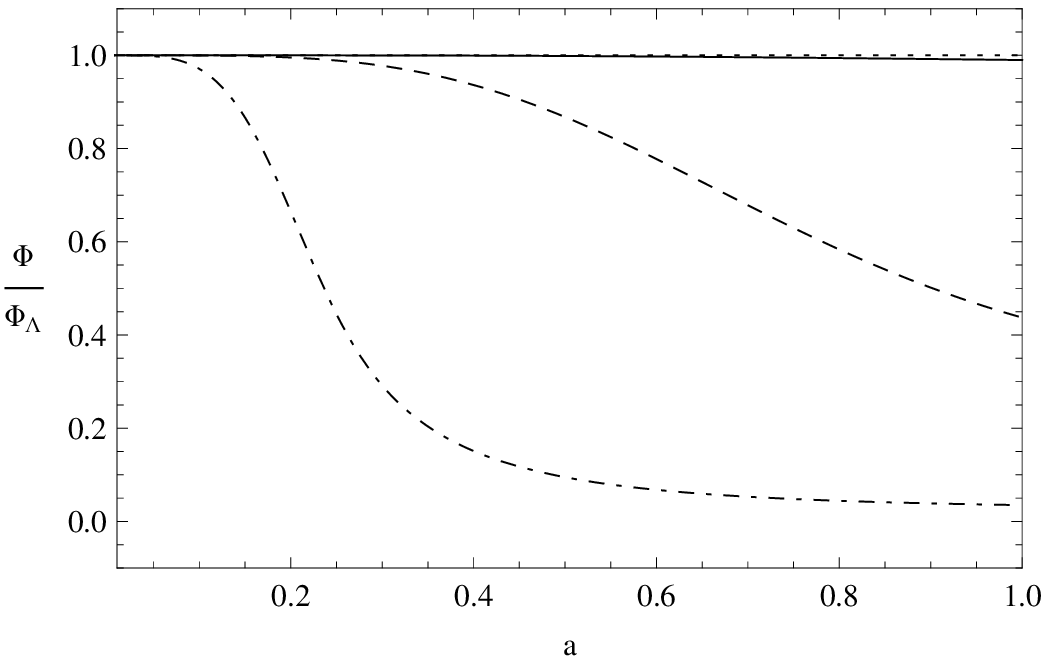}\includegraphics[width=0.5\columnwidth]{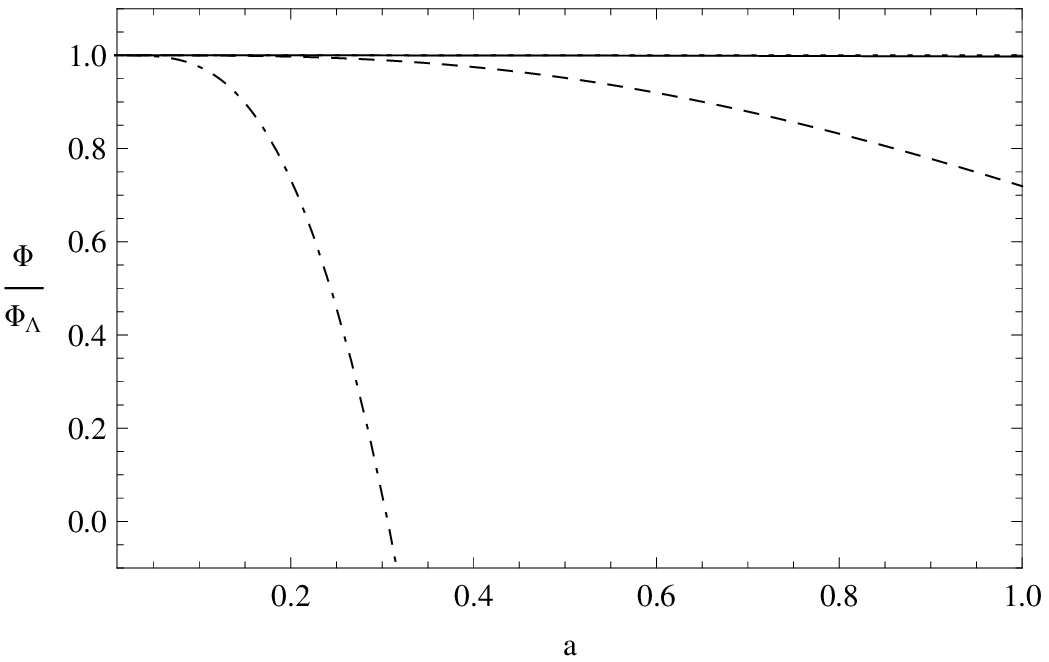}\\
\includegraphics[width=0.5\columnwidth]{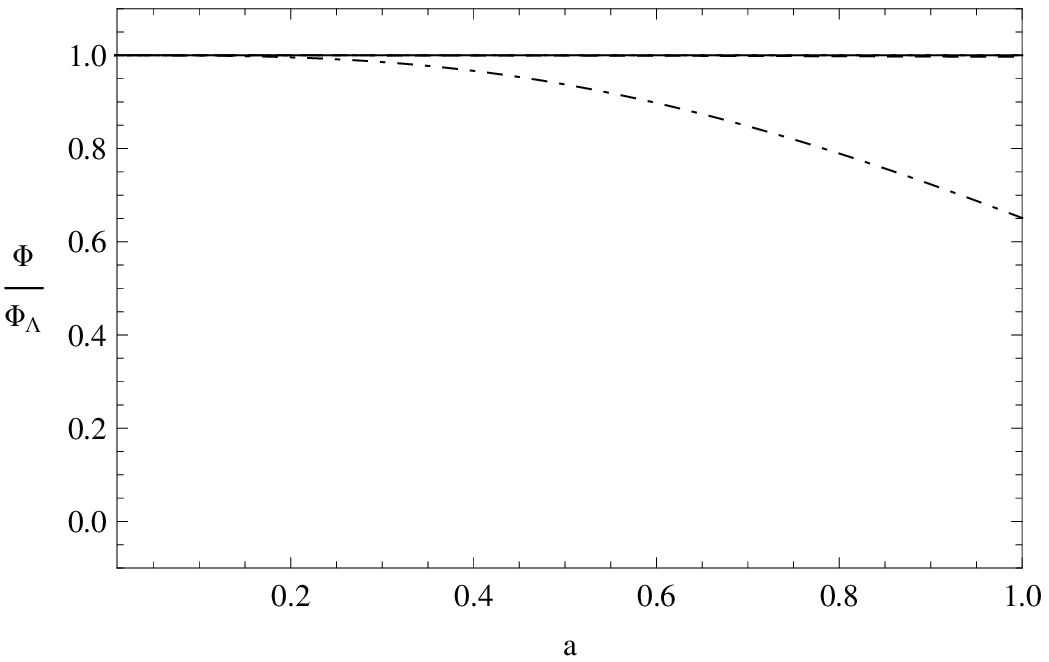}\includegraphics[width=0.5\columnwidth]{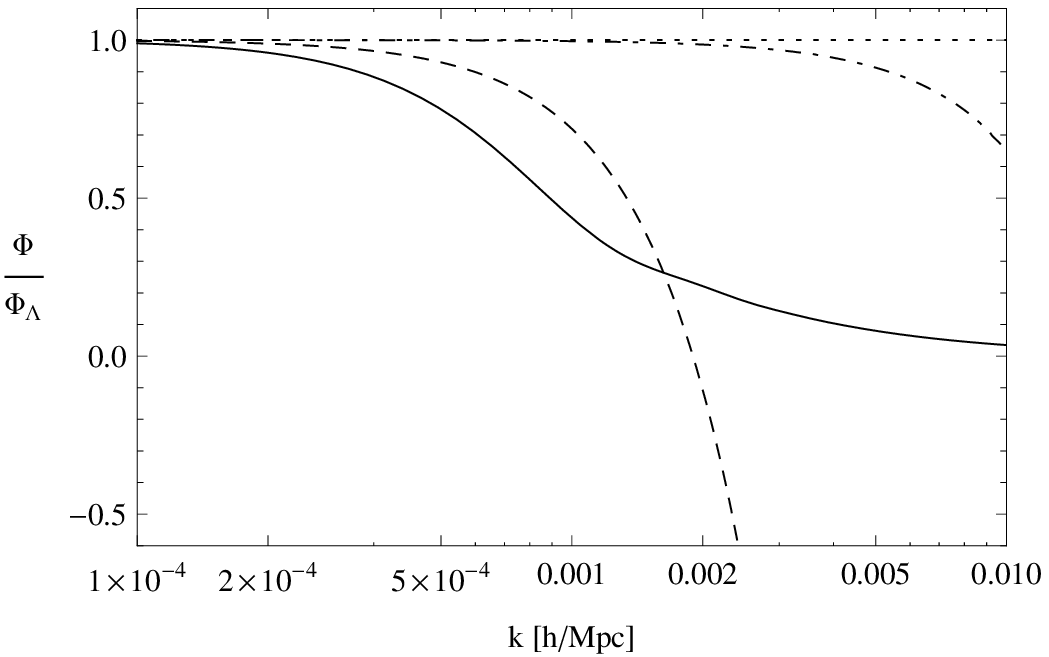}
\caption{{\bf Results for the case ${\bf n = 2}$, i.e. a $\Lambda$CDM background:} ratio $\Phi/\Phi_\Lambda$ in the truncated MIS theory for $f(\rho) = \cb^2\rho$ with $\cb^2$ constant. {\it Top-left panel:} $\cb^2 = 1$. {\it Top-right panel:} $\cb^2 = 10^{-3}$. {\it Bottom-left panel:} $\cb^2 = 10^{-8}$. In the latter three panels the profiles are functions of $a$ for $k = 0.0001, 0.001, 0.01$ $h$ Mpc$^{-1}$ fixed (solid, dashed, dot-dashed, respectively). {\it Bottom-right panel:} evolution in function of $k$ and for $a = 1$; here the solid, dashed, dot-dashed lines correspond to the cases $\cb^2 = 1, 10^{-3}, 10^{-8}$, respectively.}
\label{fig4bis}
\end{center}
\end{figure}

\begin{figure}[t]
\begin{center}
\includegraphics[width=0.5\columnwidth]{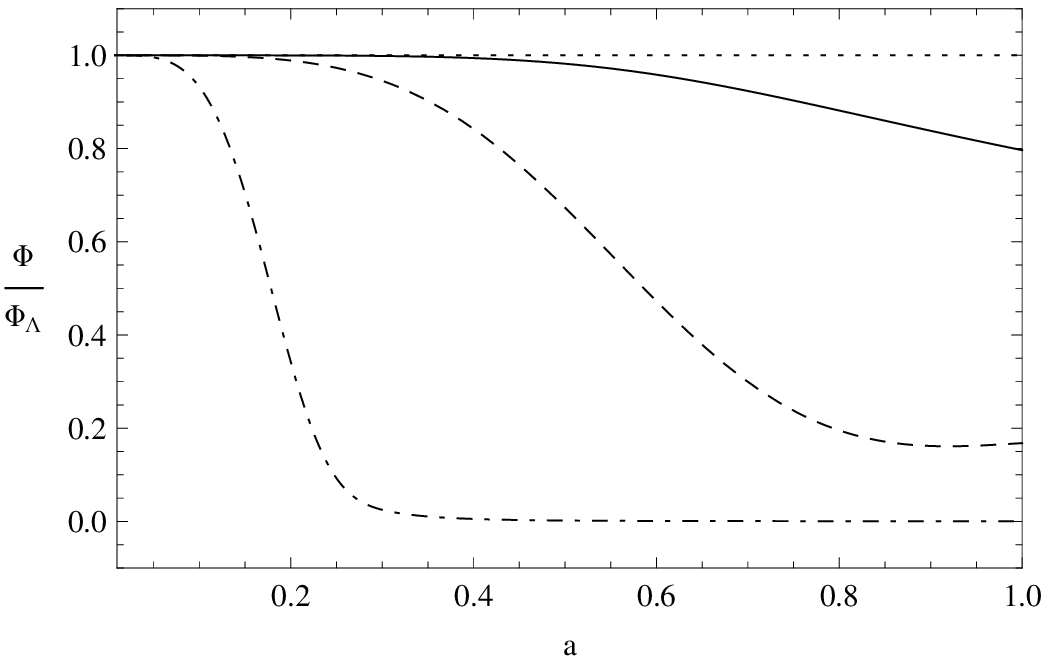}\includegraphics[width=0.5\columnwidth]{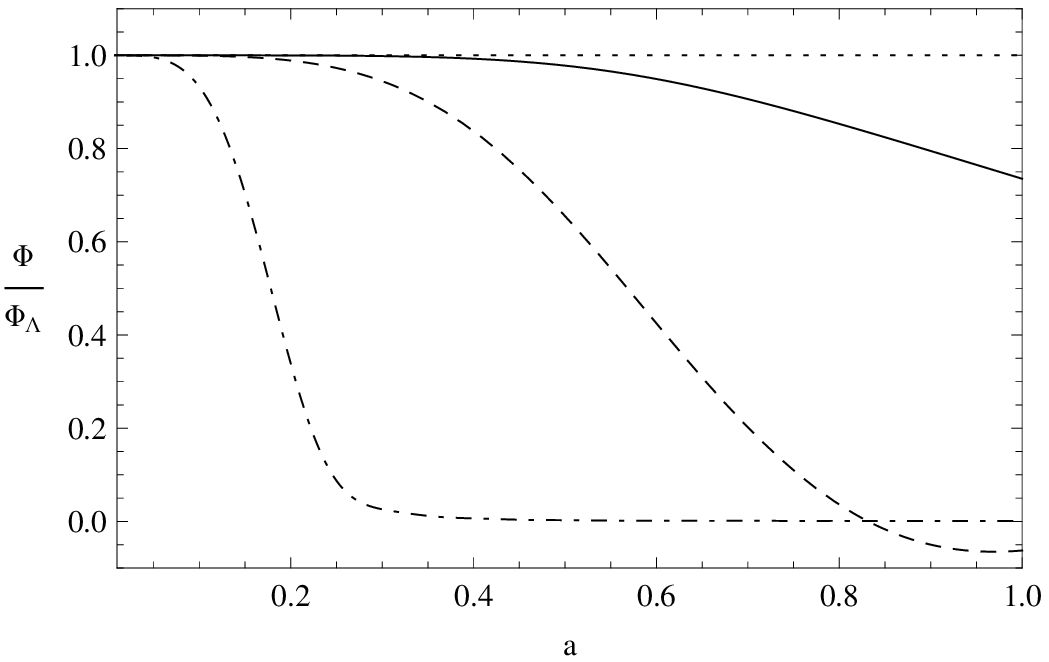}\\
\includegraphics[width=0.5\columnwidth]{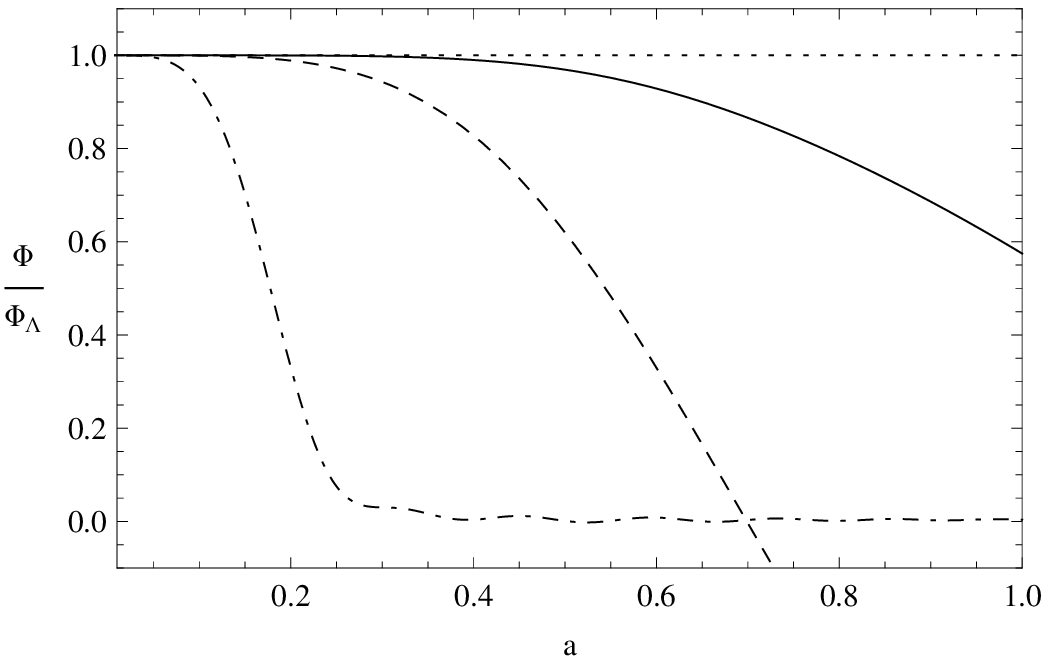}\includegraphics[width=0.5\columnwidth]{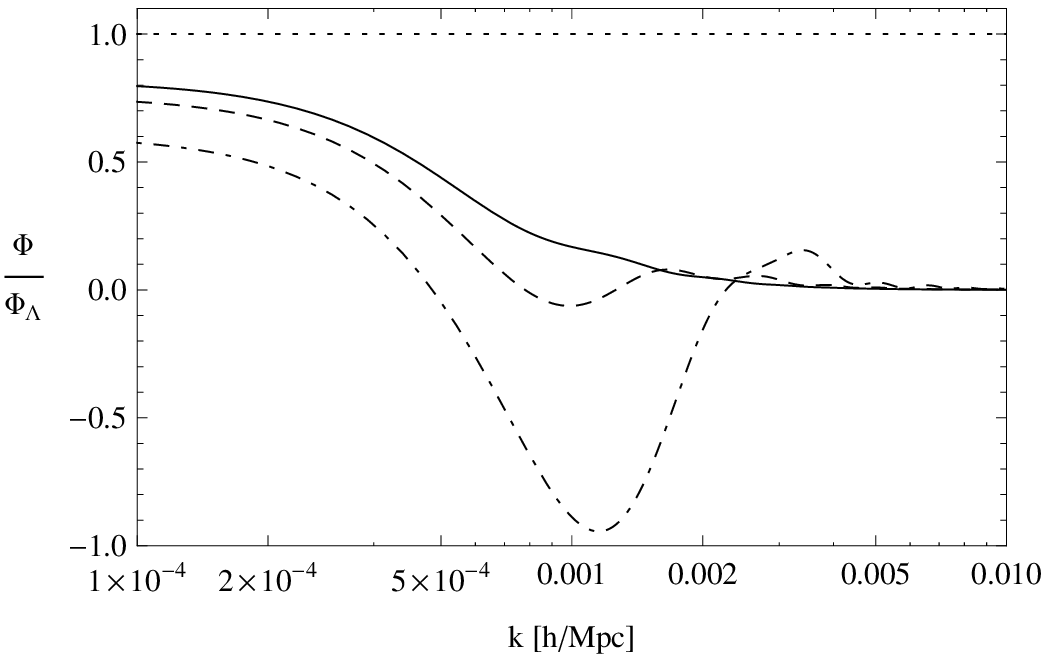}
\caption{{\bf Results for the case ${\bf n = 4}$:} ratio $\Phi/\Phi_\Lambda$ in the truncated MIS theory for $f(\rho) = \cb^2\rho$ with $\cb^2$ constant. {\it Top-left panel:} $\cb^2 = 1$. {\it Top-right panel:} $\cb^2 = 0.5$. {\it Bottom-left panel:} $\cb^2 = 0.25$. In the latter three panels the profiles are functions of $a$ for $k = 0.0001, 0.001, 0.01$ $h$ Mpc$^{-1}$ fixed (solid, dashed, dot-dashed, respectively). {\it Bottom-right panel:} evolution in function of $k$ and for $a = 1$; here the solid, dashed, dot-dashed lines correspond to the cases $\cb^2 = 1, 0.5, 0.25$, respectively.}
\label{fig4ter}
\end{center}
\end{figure}
Figs.~\ref{fig4bis} and \ref{fig4ter} show that:
\begin{enumerate}
\item For $n = 4$, lowering $\cb^2$ from unity to the limiting value $\cb^2 = 1/4$ worsens the situation.
\item  From Fig.~\ref{fig4bis} we find that for $k = 0.01$ $h$ Mpc$^{-1}$, a scale which lays at the border of the linear regime, the present value of $\Phi/\Phi_\Lambda$ is about $0.6$. This difference is compatible with the uncertainty due to the cosmic variance \cite{mukhanov2005physical}. Therefore, we may infer the qualitative constraint $\cb^2 \lesssim 10^{-8}$.
\end{enumerate}

In order to understand a value $\cb^2 \lesssim 10^{-8}$ at least roughly and tentatively, we may resort to a kinetic picture. According to a kinetic-theory based close-to-equilibrium analysis by Israel and Stewart \cite{Israel:1979wp}, the propagation speed of viscous pulses is of the order $T/m \ll 1$ for nonrelativistic particles of mass $m$, i.e.,  of the order of the adiabatic speed of sound. Assuming here the radiation temperature at matter-radiation equality $T_{\rm eq} \approx 10^{4}K \approx 0.86 \cdot 10^{-9}\mathrm{GeV}$ as an upper limit for the matter temperature, a value $\cb^2 \approx 10^{-8}$ corresponds to a lower mass limit at $0.1$ GeV. On the other hand, we must also take into account the lower limit $\cb^2 \gg 10^{-11}$ [cf. the discussion after Eq.~\eqref{taull}]. The range $10^{-11} \ll \cb^2 \lesssim 10^{-8}$ then corresponds to a mass interval $0.1 \mbox{ GeV} \lesssim m \ll 100 \mbox{ GeV}$. Note that this rough estimate is compatible with the predicted mass of DM candidates such as the WIMPs, and in particular the neutralino \cite{Bertone:2004pz}.

\newpage


\section{Conclusions}\label{sec:conclusions}

We have investigated cosmological scenarios in which the matter content of the Universe is given by a bulk viscous fluid. The bulk viscous pressure was described by Eckart's theory as well as by the causal M\"{u}ller-Israel-Stewart's theory, both in its full and truncated forms. The causal theories introduce
a relaxation time as an additional parameter into the cosmological dynamics together with a finite viscous speed of sound. Imposing a homogeneous and isotropic background dynamics with the $\Lambda$CDM model as a special case, the causal transport equations for the viscous pressure reduce to relations between the propagation speed of viscous pulses and the relaxation time. The requirement of a finite relaxation time implies upper and lower limits for the viscous sound speed. We calculated numerically the first-order gravitational potential for each of the viscous theories and compared it with the $\Lambda$CDM result.

While on the largest scales and for high redshifts all theories are similar, there are substantial and even dramatic differences on smaller scales and low redshifts. Since the gravitational potential of the $\Lambda$CDM model is supposed to provide an approximately reliable ``standard" description of the CMB power spectrum, any theory with strong deviations from this standard is very likely also disfavoured by the data. On this basis, Eckart's theory seems to be ruled out, despite of the fact that it is able to reproduce the matter power spectrum \cite{HipolitoRicaldi:2009je, HipolitoRicaldi:2010mf}. This result confirms an analysis of the integrated Sachs-Wolfe effect in \cite{Li:2009mf}. Likewise, the potentials of the full M\"{u}ller-Israel-Stewart's theory differ too strongly from its $\Lambda$CDM counterparts to be a promising alternative. The reason here is a lower limit $1/2$ for the square of the viscous propagation speed, which is much too high for a behaviour similar to that of the $\Lambda$CDM model. The only potentially valid option is the truncated causal theory for propagation speeds within the range $10^{-11} \ll \cb^2 \lesssim 10^{-8}$. In the present context, the truncated theory is not an approximation to the full theory but a phenomenological description on its own. Tentatively, values of the order of $10^{-11} \ll \cb^2 \lesssim 10^{-8}$ can be related to a nonrelativistic particle with a mass in the interval $0.1 \mbox{ GeV} < m < 100 \mbox{ GeV}$ which corresponds to a typical WIMP mass scale. A more quantitative analysis of the CMB power spectrum for the causal viscous model will be the subject of future work.


\acknowledgments

OFP was supported by the CNPq (Brazil) contract 150143/2010-9. JCF and WZ are also grateful to CNPq (Brazil) for partial financial support.

\bibliography{bibliovisc.bib}
\bibliographystyle{JHEP}

\end{document}